\documentclass[
prb,
superscriptaddress,
amsmath,
amssymb
]{revtex4}

\usepackage{graphicx}

\begin{document}

\title{Dirac-Line Criticality and Emergent Horizons in Weyl Lifshitz Transitions}

\author{Iftekher S. Chowdhury}
\affiliation{School of Mathematical and Physical Sciences, Macquarie University, Sydney, NSW 2000, Australia}

\author{Hom Nath Dhungana}
\affiliation{Charles Sturt University, Sydney, NSW, Australia}

\author{Shah Haque}
\affiliation{Southern Cross Institute, Sydney, NSW, Australia}

\author{Hind Adawi}
\affiliation{Department of Physical Sciences, Physics Division, College of Science, Jazan University, Jazan 45142, Saudi Arabia}

\author{Eric Howard}
\affiliation{School of Mathematical and Physical Sciences, Macquarie University, Sydney, NSW 2000, Australia}
\affiliation{Southern Cross Institute, Sydney, NSW, Australia}

\date{\today\\[1.8em]}

\begin{abstract}
Type-II Weyl fermions may emerge behind the event horizon of black holes. We employ the Painlev\'e-Gullstrand metric to study the surface of the Lifshitz transition at the horizon, equivalent to the interface separating the type-I and type-II Weyl states. We find several analogies between the black hole horizon and the transformation of type-I to type-II Weyl fermions through the Dirac line. We analyze the symmetry-protected topological order at the Lifshitz transition originating in semimetals. The emergence of Hawking radiation in Weyl semimetals is discussed. We show that the transition state from type-I to type-II Dirac fermions can be viewed as a black-hole horizon, which exhibits unique characteristics, including a Dirac-line Fermi surface with a nontrivial topological invariant and a critical chiral anomaly effect.

\par\vspace{0.8em}

\end{abstract}


\maketitle

\begingroup
\renewcommand\thefootnote{}
\footnotetext{
Author email addresses: Iftekher S. Chowdhury, md-iftekher.chowdhury@students.mq.edu.au; 
Hom Nath Dhungana, hdhung01@postoffice.csu.edu.au; 
Shah Haque, shah.haque@sci.edu.au; 
hindadawi@jazanu.edu.sa; 
Eric Howard, eric.howard@mq.edu.au. 
Corresponding author: Eric Howard, eric.howard@mq.edu.au.
}
\endgroup

\section{Introduction}
\label{sec:Introduction}

Topological semimetals host interesting new types of fermions as low-energy quasiparticles.
They not only exhibit novel physical properties such as distinctive topological surface states,
large linear magnetoresistance, and chiral anomaly, but also offer a versatile platform for simulating relativistic particles of high-energy physics as well as ``new particles'' that have no counterparts in high-energy physics. 
Recently, realizing new fermions, such as type-I and type-II Dirac/Weyl fermions in condensed matter systems, has attracted considerable attention. 
In relativistic theories there are several scenarios of emerging of the type-II Weyl points.  
In particular, the transition from the type-I to the type-II Weyl points occurs at the black hole event horizon.
The type-II Weyl point may also emerge as the intermediate state of the topological
Lifshitz transition, at which the Fermi surfaces exchange their global topological charge $N_3$.
This Weyl point also naturally appears if the relativistic Weyl fermions are not fundamental, but emerge in the
low energy sector of the fermionic quantum vacuum, for example, in the vacuum of the real (Majorana)
fermions.
These scenarios will be discussed here in connection to the topological materials. Some of
these considerations suggest that the inhomogeneous Weyl semimetal can serve
as a platform for simulating the black hole with stationary
metric and Hawking radiation before the equilibrium is reached. Situations
when the topological invariants are transported between the Fermi surfaces
through type II Weyl point will be considered.

Topological Dirac and Weyl semimetals are characterized with fourfold and twofold linear band crossings at the Fermi level (the so-called Dirac and Weyl points), respectively. They can be further classified into two types by fermiology. One is the class of type-I Dirac/Weyl semimetals, which has a typical conical dispersion and point-like Fermi surface. The other is the class of type-II Dirac/Weyl semimetals, which manifests in an overtilted cone-shape band structure, possessing both electron and hole pockets that contact at the type-II Dirac/Weyl point. The type-III Dirac semimetal is distinct from both type-I and type-II semimetals. A type-III Dirac point is also a protected band crossing point, but appears at the contact of a line-like Fermi surface. Unlike Fermi surfaces of other topological semimetals, such a unique line-like Fermi surface, so-called Dirac line, is protected by a topological invariant which is an integer only for the relatively rare Dirac line.

Interestingly, the type-III Dirac semimetal can also be viewed as the critical state of Lifshitz transition between type-I and type-II Dirac semimetals. The Lifshitz transition was investigated recently and a solid-state realization of black-hole-horizon analogue based on inhomogeneous topological semimetals was proposed.
It is hoped that topological semimetals will provide an alternative way to observe a black-hole horizon in condensed-matter systems. So far, however, no material system is known to be a type-III Dirac semimetal. Next, we will fill this gap by demonstrating Zn$_2$In$_2$S$_5$ to be the first type-III Dirac semimetal, and how to realize its black-hole-horizon analogue.

In the chiral gauge theory of weak interactions, the fundamental elementary particles are Weyl fermions with a pronounced asymmetry between the $SU(2)$ doublet of left-handed
Weyl fermions and the $SU(2)$ singlet of right-handed Weyl fermions. The masslessness of the Weyl fermions is topologically protected.\cite{NielsenNinomiya1981} The corresponding topological invariant
-- the Chern number -- has values $N_{3}=-1$ and $N_{3}=+1$ for the left and right particles respectively.\cite{Volovik2003}   The gapless Weyl fermions are at the origin of the anomalies in quantum field theories, such as chiral anomaly, and the corresponding symmetry protected Chern numbers characterize  the anomalous action.\cite{Volovik2003}
The Dirac particles, which emerge below
the symmetry breaking electroweak transition, are the composite objects obtained by the
doublet-singlet mixing of Weyl fermions with opposite chirality. The topological invariants $N_{3}=\pm 1$ of left and right Weyl fermions cancel each other, and without the topological and symmetry  protection the Dirac particles become massive.

Investigations in condensed matter reveal abundant and novel
physics originating from the Weyl fermionic excitations, that live in the vicinity
of the topologically protected touching point of two bands.\cite{NeumannWigner1929,Novikov1981}
Such diabolical (conical) point represents the monopole in the Berry
phase flux,\cite{Simon1983,Volovik1987} and it is described
by the same momentum-space Chern number $N_{3}$.\cite{Volovik2003}
 Weyl fermionic excitations are known to exist in the chiral
superfluid $^{3}$He-A, where the related effects \textendash{} chiral
anomaly\cite{Bevan1997,Volovik2003} and chiral magnetic effect\cite{Krusius1998,Volovik1998}
\textendash{} have been experimentally observed, and in electronic
topological semimetals.\cite{Herring1937,Abrikosov1971,Abrikosov1972,NielsenNinomiya1983,Burkov2011a,Burkov2011b,Weng2015,Huang2015,Lv2015,Xu2015,Lu2015,Hasan2017}
The Weyl points supported by the higher values of the Chern number, $|N_{3}|>1$, are also possible.\cite{VolovikKonyshev1988} In this case instead of the conical point with linear spectrum of fermions, one has the higher order band touching point, when for example the spectrum is linear in one direction and quadratic in the other directions.\cite{Volovik2003} 
These are the so-called semi-Dirac or semi-Weyl semimetals.\cite{PardoPickett2009,BanerjeePickett2012} 

Recently the attention is attracted to the type-II Weyl points.\cite{Soluyanov2015,YongXu2015,Chang2016,Autes2016,Xu2016,Jiang2016,Beenakker2016}
A remarkable property of this type of Weyl point is that it is the node of co-dimension 3 in the 3D momentum space, which is
accompanied by the nodes of the co-dimension less than three: 
the nodes of co-dimension 1 (Fermi surfaces) or nodes of co-dimension 2 (Dirac lines).
The transition between the type I and type II Weyl points is the quantum phase transition,   while the symmetry does not necessarily change  at this transition.
The quantum phase transitions with the rearrangement of the topology of the energy spectrum, at which the symmetry remains the same, are called Lifshitz transitions. Originally  I.M. Lifshitz introduced the topological transitions in metals, at which  the connectedness of
the Fermi surface changes.\cite{Lifshitz1959}  Many new types of Lifshitz transition become possible, where the topologically protected nodes of other co-dimensions are involved.\cite{Volovik2017}.
 There is a variety of topological numbers, which characterize the momentum
space manifolds of zeros. Together with the geometry of the shapes of the manifolds, this makes the Lifshitz transitions widespread in fermionic systems.

At the Lifshitz
transition between type-I and type-II fermions the Dirac lines may
also emerge, which are supported by the combined action of topology and
symmetry.
The type-II Weyl and Dirac points also emerge as the intermediate
states of the topological Lifshitz transitions.
In one case the type-II Weyl point connects
the Fermi pockets, and the Lifshitz transition corresponds to the transfer of the Berry flux between the Fermi pockets. In the other case the type-II Weyl point connects the outer and inner Fermi surfaces. At the Lifshitz transition the Weyl point is released from both Fermi surfaces. They lose their Berry flux, which guarantees the global stability, and without the topological support the inner surface disappears after shrinking to a point at the second Lifshitz transition.
For the interacting electrons, the Lifshitz transitions may lead to the formation of the dispersionless  (flat) band with zero energy and singular density of states, which opens the route to room-temperature superconductivity. Originally the idea of the enhancement of $T_c$ due to flat band has been put forward by the nuclear physics community, and this also demonstrates the close connections between different areas of physics.

We describe the transformation of the type-I to type-II Weyl fermions
through the intermediate Dirac line. Such transition may occur not only in semimetals, but also
in chiral superfluid, where the transition is regulated by superflow  due to Doppler effect
experienced by Weyl excitations. The symmetry
protected topological number of Dirac line appearing at Lifshitz
transition is discussed. We consider the behavior of 
the spectrum of Weyl fermions across the event horizon
using the Painlev\'e-Gullstrand space-time. Behind the horizon the Weyl fermions with  type II spectrum
emerge. The Fermi surfaces, which touch each other at the type-II Weyl point, become closed when the Planck scale physics is involved.
Simulation of the event horizon and Hawking 
radiation in Weyl and Dirac semimetals is discussed. We consider Lifshitz transitions, which are governed by the interplay of different topological invariants, for example, the transfer of global topological invariants between the Fermi surfaces. The formation of the flat band in the vicinity of the topological transition is considered. Finally in Sec. VI we review our results and discuss some open questions, in particular in relation to the possibility of room-temperature superconductivity in exotic topological materials.

\section{Theoretical Model and Methods}
\label{Theoretical Model and Methods}

An example of presence of the type-II Weyl fermions in relativistic theories is when
the relativistic Weyl fermions are not fundamental, but represent the fermionic excitations in the
low energy sector of the fermionic quantum vacuum.\cite{FrogNielBook, Volovik2003, Horava2005} The type-I and type-II Weyl fermions may emerge, for example, in the vacuum of the real (Majorana)
fermions.\cite{VolovikZubkov2014} The general form of the relativistic Hamiltonian for the emergent Weyl fermions is obtained by the linear expansion in the vicinity of the topologically protected Weyl
point ${\bf p}^{(0)}$ with Chern number $N_3=\pm 1$: 
\begin{equation}
H=e_{k}^{j}(p_{j}-p_{j}^{(0)})\hat{\sigma}^{k}+e_{0}^{j}(p_{j}-p_{j}^{(0)})\,.\label{HamiltonianGeneral}
\end{equation}
This expansion suggests that the position ${\bf p}^{(0)}$ of the Weyl
point, when it depends on coordinates, serves as the $U(1)$ gauge field, ${\bf A}({\bf r},t)\equiv {\bf p}^{(0)}({\bf r},t)$, acting on relativistic fermions. The parameters $e_{k}^{j}({\bf r},t)$
and $e_{0}^{j}({\bf r},t)$ play the role of the emergent tetrad fields, describing the gravity experienced by Weyl fermions. 

The energy spectrum of the Weyl fermions depends on the ratio between
the two terms in the rhs of Eq.(\ref{HamiltonianGeneral} ), i.e.
on the parameter $|e_{0}^{j}[e^{-1}]_{j}^{k}|$.\cite{VolovikZubkov2014}
When $|e_{0}^{j}[e^{-1}]_{j}^{k}|<1$ one has the conventional Weyl point. The Weyl cone is tilted, if $e_{0}^{j}\neq 0$. At $|e_{0}^{j}[e^{-1}]_{j}^{k}|>1$ the cone is overtilted, and
two Fermi surfaces appear, which touch each other at the Weyl point.
In condensed matter this regime is called the type-II Weyl, as distinct from the conventional Weyl point, which is called type-I.\cite{Soluyanov2015} The Lifshitz transition between the two regimes occurs at $|e_{0}^{j}[e^{-1}]_{j}^{k}|=1$. In the relativistic regime, the spectrum of Weyl fermions at the transition contains zeros of co-dimension 2 -- 
the Dirac line. In general, the existence of the nodal lines requires the special symmetry: they are protected by topology in combination with symmetry. 

There are indications that in some materials the maximum of the superconducting transition temperature occurs just in the vicinity of the Lifshitz transitions (see also the flat-band discussion below). In particular, the enhancement of $T_c$ at the type-I to-
type-II topological transition in Weyl semimetals has been discussed in Ref.\cite{ShapiroShapiro2017}.

To reveal properties of this Lifshitz transition,
let us start with considering the topological charge of the nodal
line using a simple choice of the tetrads for the relativistic Weyl fermions in the gravitational field: 
\begin{equation}
H=c{\mbox{\boldmath\ensuremath{\sigma}}}\cdot\hat{{\bf p}}-fcp_{z}\,.
\label{HamiltonianSimple}
\end{equation}
For $f\neq0$ the Weyl cone is tilted, and for $f>1$ the type-II
Weyl point takes place when the tilted Weyl cone crosses zero energy.
At the boundary between the two regimes, with $f=1$, the Hamiltonian has the form
\begin{equation}
H=\begin{pmatrix}0 & c(p_x+ ip_y)\\
c(p_x- ip_y) & -2cp_z
\end{pmatrix}\,,\label{eq:22Matrix}
\end{equation}
and the energy spectrum has the
nodal line on the $p_{z}$-axis, i.e.  $E({\bf p}_{\perp}=0,p_z)=0$ for all $p_z$.
We consider several approaches to characterize stability of the nodal Dirac lines in relativistic systems,
which could be extended to condensed matter systems.

In the first approach we take into account that the matrix in Eq.(\ref{eq:22Matrix}) belongs to the class of
the $2n\times2n$ matrices of the type: 
\begin{equation}
H=\begin{pmatrix}0 & B({\bf p})\\
B^{+}({\bf p}) & C({\bf p})
\end{pmatrix}\,,\label{eq:Matrix}
\end{equation}
and the topological properties of the considered nodes in the spectrum are characteristics of this class.
Of course, it is difficult to expect such matrices in real physical
systems, except for the case of $n=1$, which naturally emerges at Lifshitz transition. But
it is instructive to consider the general $n$ case. The determinant
of such matrix is the product of the determinants of matrices $B$
and $B^{+}$: 
\begin{equation}
D(H)=-D(B)D^{*}(B)\,.\label{eq:Determinant}
\end{equation}
The nodal lines \textendash{} zeros of co-dimension 2 \textendash{}
are zeros of $D(B)$ and are described by the winding number of the phase
$\Phi$ of the determinant $D(B)=|D(B)|e^{i\Phi}$: 
\begin{equation}
N_{2}=\oint_{C}\frac{dl}{2\pi i}D^{-1}(B)\partial_{l}D(B)={\bf tr}\oint_{C}\frac{dl}{2\pi i}B^{-1}({\bf p})\partial_{l}B({\bf p})\,,\label{eq:N2B}
\end{equation}
where $C$ is the closed loop in momentum space around the line. 
The line in momentum space with the non-zero winding number of the phase $\Phi$ is the momentum-space analog of the vortex line in superfluids and superconductors, which is characterized by the winding number of the phase of the order parameter.

For the particular case of $2\times2$ matrix in Eq.(\ref{eq:22Matrix}),
where $D(B)=B=c(p_{X}+ip_{y})$, the invariant can be written as 
\begin{equation}
N_{2}={\bf tr}\oint_{C}\frac{dl}{4\pi i}\cdot[\sigma_{z}H_{f=1}^{-1}({\bf p})\partial_{l}H_{f=1}({\bf p})]\,,\label{eq:N2}
\end{equation}
where the Dirac line corresponds to the $p_{z}$-axis. 

The form $(\ref{eq:N2})$ of invariant $N_2$
is somewhat counterintuitive, since the integral of this type represents the true integer-valued invariant only if $\sigma_{z}$ commutes or anticommutes with the Hamiltonian. The latter  does not happen here, nevertheless the integral is still integer-valued, which can be shown in a
straightforward way. For $p_{z}=0$ the Hamiltonian anticommutes with
$\sigma_{z}$, and the integral is the well defined topological invariant
with $N_{2}=1$ for any $f$. At $p_{z}\neq0$ the Hamiltonian does
not anticommute with $\sigma_{z}$. However, Eq.(\ref{eq:N2}) remains
integer for the general $p_{z}$ if $f=1$. 

To see that we apply the second approach. Let us consider
$p_{z}$ as parameter and the arbitrary loops around the line ${\bf p}_{\perp}=0$
with fixed $p_{z}$. Taking into account that 
\begin{equation}
H^{-1}(f=1)=\frac{1}{p_{\perp}^{2}}\left(c{\mbox{\boldmath\ensuremath{\sigma}}}\cdot\hat{{\bf p}}+cp_{z}\right)\,,\label{InverseHamiltonian}
\end{equation}
one obtains that the variation of $N_2$ over $p_z$ is zero:
\begin{equation}
\frac{dN_{2}(p_{z})}{dp_{z}}=0\,\,,\,\,N_{2}(p_{z})={\bf tr}\oint_{C(p_{z})}\frac{dl}{4\pi i}\cdot[\sigma_{z}H_{f=1}^{-1}({\bf p}_{\perp},p_{z})\partial_{l}H_{f=1}({\bf p}_{\perp},p_{z})]\,.\label{eq:N2derivative}
\end{equation}
Thus the integral $N_{2}(p_{z})=1$ for any $p_{z}$ at $f=1$.

The stability of the vortex line in momentum space can be understood through consideration in
terms of the determinant of the Hamiltonian matrix, i.e. $D(H)$,
in a way somewhat similar to that in Ref.\cite{KimmeHyart2016} 
\begin{equation}
D(H)=c^{2}p_{z}^{2}(f^{2}-1)-c^{2}p_{\perp}^{2}\,.\label{eq:Drel}
\end{equation}
$D(H)$ is nonzero for $0<f<1$, is zero on line at $f=1$,
and has zeros on the conical Fermi surface at $f>1$. For $f=1$ one
can define the generalized root $q(H)$ of det $H$ \textendash{}
a polynomial function of the matrix elements of the Hamiltonian \textendash{}
in such a way that $|q(H)|^{2}=|D(H)|$. So $q(H)$ is our $D(B)$
in Eq.(\ref{eq:Determinant}). The corresponding polynomial is 
\begin{equation}
q(H_{f=1})=D(B)=c(p_{x}+ip_{y})\,.\label{eq:qH}
\end{equation}
It has zero on the line ${\bf p}_{\perp}=0$ which is protected by
$2\pi$ winding of the phase of $q$ around the line. This gives rise
to the topologically stable zero in the determinant $D(H)$ and thus to
topologically stable zero in the quasiparticle spectrum. For $f\neq 1$,
the integral Eq.(\ref{eq:N2}) depends on $p_{z}$ and on the radius
of the closed loop $C$.


This type of transition can be seen on example of the modification of the model describing the rhombohedral graphite
\cite{HeikkilaVolovik2015}, with
\begin{equation}
B= (p_x +ip_y)\left( p_x +ip_y + t_+e^{ip_z a} +  t_-e^{-ip_z a}  \right)\,.
\label{eq:CrossingLines}
\end{equation}
This model has two nodal lines -- the straight one along the $z$ direction and the spiral around the straight one. Due to the lattice periodicity originating from the layers type construction along $z$ direction, the spectrum of quasiparticles along $p_z$ direction can be described in terms of the one dimensional Brillouin zone. As a result the nodes in the spectrum are the closed loops. From the viewpoint of knot theory, these two nodal lines form a Hopf link --  the simplest nontrivial link consisting of two unknots.\cite{Kauffman2001}. The Hopf link with $t_+>t_-$ is the mirror image of the one with $t_+<t_-$, and these two configurations cannot be connected with combination of Reidemeister moves.\cite{Kauffman2001} This means that they are not ambient isotopic, and they may transform to each other only via the special type of Lifshitz transition, which in our case occurs at $t_+=t_-$. To characterize the difference between these kinds of two Hopf linked nodal lines, we assign a fixed direction and calculate the linking number via $N_{l}=(\sum_{p} {\epsilon_{p}} )/2$, where $p$ is the crossing in the diagram of Hopf link of nodal lines, and $\epsilon_{p}$ is the sign of the oriented crossing.\cite{Kauffman2001} From the corresponding nodal-line figure, we can find that $N_{l}=1$ for $t_+>t_-$, while $N_{l}=-1$ for $t_+<t_-$ respectively.

On other examples of knotted nodal lines see e.g. in Ref.\cite{Nodal-knot2017}.

In general relativity the convenient stationary metric for the black hole both
outside and inside the horizon is provided in the Painlev\'e-Gullstrand
spacetime\cite{Painleve} with the line element:
\begin{equation}
ds^{2}=-c^{2}dt^{2}+(d{\bf r}-{\bf v}dt)^{2}=-(c^{2}-v^{2})dt^{2}-2{\bf v}d{\bf r}dt+d{\bf r}^{2}\,.\label{Painleve}
\end{equation}
This is stationary but not static metric, which is expressed in terms
of the velocity field ${\bf v}({\bf r})$ describing the frame drag in the
gravitational field. The Painlev\'e-Gullstrand is equivalent to the so-called acoustic metric, 
\cite{unruh1,unruh2,Kraus1994}
where ${\bf v}({\bf r})$ is the velocity of the normal  or superfluid liquid.

We shall see that  analogous transition occurs for the relativistic fermions when the event horizon of the black hole is crossed and the frame drag velocity exceeds the speed of light.  This analogy suggests a route for simulation of an event horizon in inhomogeneous condensed matter systems, which is accompanied by the analog of Hawking radiation. 

For the spherical black hole the frame drag velocity field (the velocity of the free-falling observer) is radial: 
\begin{equation}
{\bf v}({\bf r})=-\hat{{\bf r}}c\sqrt{\frac{r_{h}}{r}}~,~r_{h}=\frac{2MG}{c^{2}}\,.\label{VelocityField}
\end{equation}
Here $M$ is the mass of the black hole; $r_{h}$ is the radius of
the horizon; $G$ is the Newton gravitational constant. The minus
sign in Eq.(\ref{VelocityField}) gives the metric in case of the black hole,
while the plus sign would characterize the gravity of a white hole. 

Let us consider a Weyl particle in the Painlev\'e-Gullstrand
space-time. The 
tetrad field corresponding to the metric in Eq.(\ref{Painleve}) has the form:\cite{Doran2000} 
\begin{equation}
e_{k}^{j}=c\delta_{k}^{j}\ \text{and}\ e_{0}^{j}=v^{j}  \,,
\label{PGTetrad}
\end{equation}
which leads to the following Hamiltonian:
\begin{equation}
H=\pm c{\mbox{\boldmath\ensuremath{\sigma}}}\cdot{\bf p}-p_{r}v(r)+\frac{c^{2}p^{2}}{E_{{\rm UV}}}\,\,,\,\,v(r)=c\sqrt{\frac{r_{h}}{r}}\,.\label{HamiltonianBH}
\end{equation}
Here the plus and minus signs correspond to the right handed and left
handed fermions respectively; $p_{r}$ is the radial momentum of fermions.
The second term in the rhs of (\ref{HamiltonianBH}) is the Doppler
shift ${\bf p}\cdot{\bf v}({\bf r})$ caused by the frame drag velocity (compare
with the analogous chiral-superfluid Hamiltonian for chiral superfluid). 

The third
term in Eq.(\ref{HamiltonianBH}) is the added nonlinear dispersion to take into account the Planckian physics,
which becomes important inside the horizon. The parameter
$E_{{\rm UV}}$ in the third term is the ultraviolet (UV) energy scale, at which the Lorentz invariance
is violated. The UV scale is typically associated with but does not necessarily correspond to the Planck energy scale.\cite{Kostelecky2011,Rubtsov2016}
 For the interacting fermions such term can arise in effective
Hamiltonian $H=G^{-1}(\omega=0,{\bf p})$ even without violation of
Lorentz invariance on the fundamental level:\cite{Volovik2010} the Green's
function $G(\omega,{\bf p})$ may still be relativistic invariant,
while the Lorentz invariance of the Hamiltonian is violated due to
the existence of the heat bath reference frame. In this case the UV scale is below the Planck scale.

Behind the black hole horizon,
the Weyl point in the spectrum transforms to the pair of the closed Fermi
surfaces with the touching point: the type-II Weyl point. The $p^{2}$
term in Eq.(\ref{HamiltonianBH}) makes the Fermi surfaces attached
to the type-II Weyl point closed, while it provides only a small
correction when $cp\ll E_{{\rm UV}}$. The latter is valid 
if the spectrum inside the black hole is considered in the vicinity of the horizon,
where $r_{h}-r\ll r_{h}$, see  the corresponding Fermi-surface plots,
at positions $r=0.95r_{h}$ and $r=0.9r_{h}$.
 Similar to the situation in section II, near
the horizon the Dirac(Weyl) cone is tilted, and behind the horizon
it crosses zero energy and forms the Fermi surfaces corresponding
to the type-II Weyl points. But there is no  Dirac line at the
horizon, where Lifshitz transition occurs: as we mentioned before, this is because of the quadratic term, which violates the Lorentz invariance. 
Instead, the Fermi pockets start to grow from the Weyl point  with $|p_{r}|<p_{{\rm UV}}(v(r)-c)/c\ll p_{{\rm UV}}$,
when the horizon is crossed.

In the full equilibrium the Fermi pockets must be occupied by particles and "holes".
One of the mechanisms of the filling of the Fermi pockets in the process of equilibration will be observed by external observer as Hawking radiation. The Hawking temperature is determined by effective gravitational field at the horizon:
\begin{equation}
T_{{\rm H}}=\frac{\hbar}{2\pi}\left(\frac{dv}{dr}\right)_{r=r_{h}}\,
\label{HawkingT}
\end{equation}
If the Hawking radiation is the dominating process of the black hole evaporation, the lifetime of the black hole
is astronomical. However, the other much faster mechanisms involving the trans-Planckian physics are not excluded.


Based on the previous discussions,
one can suggest a new route through which the black hole horizon and ergosurface
can be simulated using the inhomogeneous condensed
systems with emergent type-I and type-II Weyl fermions. The interface
which separates the regions of type-I and type-II Weyl points may serve as the event horizon, on which 
the Lifshitz
transition takes place. In general case, such an artificial horizon may
have the shape different from the spherical surface. The shape of the horizon is not important
if we are interested in the local temperature 
of Hawking radiation, which is determined by the local effective gravity at the horizon.

Let us consider the completely flat artificial event horizon on example of Eq.(\ref{HamiltonianSimple}). We assume that the parameter $f$ depends on $z$, and $f(z)$ crosses unity at $z=z_{\rm hor}$.
  The plane $z=z_{\rm hor}$ separates  the region with type-I Weyl fermions ($f(z)<1$) from the region with type-II Weyl 
fermions  ($f(z)>1$).  This plane corresponds to the event horizon, while the ergoplane 
can be obtained for the other orientations of the plane with respect to the axis $z$ (review on
artificial horizons and ergoregions in acoustic metric see in Ref. \onlinecite{Visser1998}). 
the corresponding schematic demonstrates the Weyl cones in the energy-momentum space ({\it bottom}) and the analogues of the light cone ({\it top})  for quasiparticles on two sides of the event horizon.
Behind the horizon the Weyl cone is overtilted so that the upper cone crosses the zero energy level,
and the Fermi surfaces (Fermi pockets) are formed, which are connected by the type-II Weyl point.
Correspondingly the future light cone is overtilted behind the horizon so that all the paths fall further into the black hole region. 

The filling of the originally empty states inside the Fermi surfaces causes the Hawking radiation.
 For the flat horizon the Hawking temperature determined by the effective gravitational field at the horizon is:
\begin{equation}
T_{{\rm H}}=\frac{\hbar c}{2\pi}\left(\frac{df}{dz}\right)_{z=z_{\rm hor}}\,
\label{HawkingTflat}
\end{equation}

Note that in the Weyl semimetals the mechanism of formation of the artificial event horizon (and its behavior
after formation) is different from the traditional mechanism, which is based on the supercritical flow of the liquid or
Bose-superfluids.\cite{unruh1,unruh2,Visser1998,Volovik2003} 
In the latter case the effective metric (the so-called acoustic metric) is produced by the flow of the liquid and
thus represents the non-static state. Due to the dissipation (caused, say, by the analogue of Hawking radiation) the flow relaxes and
reaches the sub-critical level, below which the horizon disappears. On the contrary, in semimetals the tilting of the Weyl cone occurs without
the flow of the electronic liquid, and thus the state with the horizon is fully static. The dissipation after the formation
of the horizon (caused, say, by analogue of Hawking radiation) leads to the filling of the electron and hole Fermi pockets. After the Fermi pockets are fully occupied the final state is reached, but it still contains the event horizon, though the Hawking radiation is absent. Similar
mechanism takes place in the fermionic superfluids, such as superfluid $^3$He, where depending on the parameters of the system the flow may or may not remain supercritical after the Fermi pockets are occupied, see Fig. 26.1 in Ref.\onlinecite{Volovik2003}.



Here we discuss the case, when the type-II Weyl and type-II Dirac points emerge during the Fermi surface Lifshitz transitions. By the type-II Weyl point the topological invariant $N_3$ is transported
between the Fermi surfaces.
With the multiplicity of topological
invariants for the manifolds of nodes in the fermionic spectrum, Lifshitz transitions become diverse and complex. This can be seen on examples of the Lifshitz transitions with the reconstruction of the Fermi surfaces, where several topological invariants may interplay. 

Topological invariants which are involved in the complex topological Lifshitz
transitions are: (i) the invariant $N_{1}$, which is responsible for
the local stability of the Fermi surface;\cite{Volovik2003}
(ii) the invariant $N_{3}$, which is the global invariant describing the
closed Fermi surface: when the Fermi surfaces collapse to a point, it becomes the type-I Weyl point
with the topological charge $N_{3}$; and 
(iii) the $N_{2}$ invariant in Eq.(\ref{eq:N2}) 
which characterizes the Dirac line.
All three topological invariants are involved in the complex Lifshitz transition.
 For Fermi surfaces with non-vanishing $N_{3}$, there is the type-II point attached to the Fermi surfaces
at the critical point of Lifshitz transition.  This type-II point 
has also the nontrivial $N_{2}$, with the contour $C$ chosen as the  infinitesimal loop around the cone, see also reference
{[}22{]}.\cite{YongXu2015} This is the consequence of the $\pi$ Berry phase along the  infinitesimal loop 
around the Weyl point. And of course, the invariant $N_1$ supports the local stability of the Fermi surface and does not allow to make a hole in the Fermi surface and disrupt it.

Here, we present three models, each with its own characteristics, which
exhibit complex topological Lifshitz transition induced by the interplay between
$N_{1}$, $N_{3}$ and $N_{2}$ invariants. 

The Hamiltonian for a massive Dirac particle with mass
$M$ has the form: 
\begin{equation}
H=\begin{pmatrix} {\mbox{\boldmath$\sigma$}} \cdot ( c\,{\bf p}-{\bf b}) - b_0  &M
 \\ M & -{\mbox{\boldmath$\sigma$}} \cdot ( c\,{\bf p}+{\bf b})  +b_0 \end{pmatrix} \,.
\label{eq:Fermipoint}
\end{equation}
Here the 4-vector $b_{\mu}=(b_{0},{\bf b})$ causes the shift of the positions of the
Berry phase monopoles in opposite direction and formation of two Fermi
surfaces with the global charge $N_{3}=\pm1$ if ${\bf b}^{2}>b_{0}^{2}+M^{2}$ as is
shown in the corresponding type-II Dirac schematic. One Fermi surface enclosing the
Berry phase monopole with topological charge $N_{3}=+1$ is formed
by the right-handed Weyl fermions; while the other one, which encloses
the Berry phase monopole with topological charge $N_{3}=-1$, comes
from the left-handed Weyl fermions. Positions of the monopoles
are at
\begin{equation}
{\bf p}_\pm = \pm{\bf b} \left(\frac{{\bf b}^2-b_0^2-M^2}{{\bf b}^2-b_0^2}\right)^{1/2} \,.
\label{Monopoles}
\end{equation}

At critical point of Lifshitz transition,
${\bf b}^{2}=b_{0}^{2}+M^{2}$,  two Berry phase
monopoles with opposite chirality merge forming the Dirac point with trivial topological charge
$N_{3}=0$. In contrast, the non-vanishing $N_{1}$
locally preserves Fermi surfaces at this critical
point. As a result of this interplay between $N_{3}$ and $N_{1}$,
the Fermi surfaces are attached to the Dirac point forming the type-II 
Dirac point, see the corresponding type-II Dirac schematic.
Note that as different from the Weyl point, the Dirac point  is marginal: it has  trivial global topology and is not stable, if there is no special symmetry
which can stabilize the node.\cite{Volovik2003}
Here  the type-II Dirac point appears exactly at the Lifshitz transition, similar to the 
appearance of the Dirac nodal line in the analogous nodal-line schematic discussed earlier.



For the type of transition discussed in the preceding Dirac-point discussion, the topological index $N_{2}$ is not involved, because the intermediate state is the Dirac point. 
To obtain  the type-II fermions with non-vanishing $N_{2}$ at the Lifshitz
transition, one should consider the system in which 
the Fermi surfaces are connected not by the type-II marginal Dirac point but by the
topologically stable type-II Weyl point.

The corresponding Hamiltonian is obtained by the natural extension of  the analogous chiral-superfluid Hamiltonian  for quasiparticles in chiral superfluid $^3$He-A in the presence of the superfluid current. If the current is chosen perpendicular to the directions towards the Weyl nodes, then at $v>c$ we obtain two type-II Weyl points, which connect two banana shape Fermi surfaces in the corresponding type-II Weyl schematic. 
These type-II Weyl points, in addition to the Berry monopole invariant $N_3=\pm 1$, have the 
nonzero value of the topological charge
 $|N_{2}|$ calculated for the infinitesimal closed loop $C$ around the cones. If the closed loop $C$
 is within the symmetry plane of two
Fermi surfaces, the integral is independent on the shape and the radius of the contour $C$ of
 integration. In general, however, when the symmetry is violated, only the integration over the infinitesimal 
loop gives the integer value of the invariant, see the case in which the monopole leaves the Fermi surface.

Let us now change the direction of the current.
If $\theta$ is the angle between the current and the directions to the nodes, the Hamiltonian in the laboratory
frame becomes:
\begin{equation}
H=p_{x}v+\tau_{3}\frac{p^{2}-p_{F}^{2}}{2m}+\tau_{1}c(p_{x}\sin\theta-p_{z}\cos\theta)+\tau_{2}cp_{y}\,.\label{TypeIIWeyl1}
\end{equation}
Eq.(\ref{TypeIIWeyl1}) is identical to the analogous chiral-superfluid Hamiltonian when $\theta=\pi/2$.
the corresponding type-II Weyl schematic  demonstrates the Lifshitz
transition induced by the change of $\theta$, from $\theta<\pi/2$
to $\theta>\pi/2$, at $v>c$.  When one continuously
changes the angle $\theta$ across $\theta=\pi/2$, the Berry phase monopoles
with topological charges $N_3=\pm 1$   
move counter clockwise on a Fermi sphere with radius $|\mathbf{p}|=p_{F}$
in the $p_{y}=0$ plane.   At the same time, the non-vanishing local stability
invariant $N_{1}$ with $v>c$ protects the Fermi surfaces during this process.
As a result, the Berry phase monopoles are transported between the two
Fermi surfaces, with the type-II Weyl points emerging in the intermediate state of
this topological Lifshitz transition.
Similar phenomenon with the interplay between topological invariants
$N_{1}$, $N_{2}$  and $N_{3}$ may take place in $bcc$ Fe,
see details in Ref. \onlinecite{GosalbezMartinez2015}.  

Let us consider another class of emergent type-II Weyl point, in which the Berry
phase monopole is transported across the Fermi surface. It can be represented by the following
Hamiltonian:
\begin{equation}
H=c{\mbox{\boldmath\ensuremath{\sigma}}}\cdot({\bf p}-{\bf p}^{(0)})+\frac{p^{2}-p_{F}^{2}}{2m}\,
\label{ExampleHamiltonian}
\end{equation}

In the corresponding second type-II Weyl schematic, we plot the Lifshitz transitions and the
evolution of configuration of Berry  monopole in momentum space
driven by the change of the position $\mathbf{p}^{(0)}$ of the Weyl point. The regime with $p_F > mc$ is considered.
For $|\mathbf{p}^{(0)}| < p_F$ we have two Fermi surfaces, one inside the other, but both embracing
the Weyl point with $N_3=1$, the Berry phase monopole. At the Lifshitz transition, which occurs  at  
$|\mathbf{p}^{(0)}| =p_F$, the inner and outer Fermi surfaces touch each other at the Weyl point, which becomes the peculiar type-II point. As distinct from the conventional type-II Weyl point, which connects two Fermi pockets,  this Weyl point connects the inner and outer Fermi surfaces. After the Lifshitz transition, at  $|\mathbf{p}^{(0)}| >p_F$, the Weyl point leaves both Fermi surfaces. The Fermi surfaces are again one inside the other, but 
both without the Berry flux. Finally at the second Lifshitz transition, at $|\mathbf{p}^{(0)}| =(m^2c^2 + p_F^2)/2mc$,  the inner Fermi surface collapses to the point and disappears, since the point is no more supported by the topological invariant $N_3$.

At the first Lifshitz transition, the cone formed at the touching point is again characterized by the topological invariant $N_2=1$, where 
the integral is over the infinitesimal path around the cone.

For the interacting fermions more types of Lifshitz transitions are possible -- the transitions  which involve the Weyl points of type-III and type-IV.\cite{NissinenVolovik2017} 
The interaction also leads to the formation of the flat band in the
energy spectrum --  the so-called Khodel-Shaginyan fermion 
condensate.\cite{KhodelShaginyan1990,Volovik1991,Nozieres1992}
The dispersionless energy spectrum has a singular density of states. As a result,  instead of the exponential suppression of the superconducting transition temperature $T_c$ (and of the gap $\Delta$)  in the normal metal, the flat band provides $T_c$ and $\Delta$ being proportional to the coupling constant $g$ in the Cooper channel:
\begin{equation}
\Delta_{\rm normal} = E_0  \exp\left(- \frac{1}{ g N_F}\right) \,\,\,,
 \,\,\Delta_{\rm flat\, band}  = \frac{ gV_d}{2(2\pi \hbar)^d}\,.
\label{expVSlin}
\end{equation}
Here $N_F$ is the density of states in normal metal; $d$ is the dimension of the metal; and $V_d$ is the volume of the flat band. For  nuclear systems, i.e. for $d=0$,  
the linear dependence of the gap on the coupling constant has been found by Belyaev.\cite{Belyaev1961}
The enhancement of $T_c$  in materials with the flat band opens the route to room temperature superconductivity, see review Ref.\cite{HeikkilaVolovik2016}.

The band flattening caused by electron-electron interaction in metals is the manifestation of the general phenomenon of the energy level merging due to electron-electron interaction. This effect has been recently
suggested \cite{Dolgopolov2014} to be responsible for merging  of the discrete energy levels in two-dimensional electron system in quantizing magnetic fields. 

 According to Ref.~\cite{Yudin2014} the favorable condition for the formation of such flat band is when the van Hove singularity comes close to the Fermi surface, i.e. the system is close to the Lifshitz transition  (see also Ref.~\cite{Volovik1994} for the simple Landau type model of the formation of such flat band). It is also possible that this effect is responsible for the occurrence of superconductivity with high $T$ observed in the pressurized sulfur hydride \cite{Drozdov2014,Drozdov2015}.
There is some theoretical evidence \cite{Pickett2015,Bianconi2015} that high-$T_c$ superconductivity takes place at such pressure, when the system is near the Lifshitz transition.
That is why it is not excluded that  the Khodel-Shaginyan flat band is formed  in sulfur hydride at pressure 180-200 GPa giving rise to high-T$_c$ superconductivity. The topological Lifshitz transitions with participation of the Weyl and Dirac points and Dirac lines may also lead to the formation of the flat bands in the vicinity of transitions, and thus to the enhanced $T_c$.

Here we consider the flat bands, which appear near the Lifshitz transition  
in Eq. (\ref{ExampleHamiltonian}). The arrangement of the flat bands experiences its own Lifshitz transitions in the corresponding flat-band configuration.



The energy functional of interacting loosing Weyl model is:
\begin{equation}
E[n_{1}(\mathbf{p}),n_{2}(\mathbf{p})]=\sum _{\mathbf{p}}{{\epsilon}_{1 }^{0}{n}_{1}(\mathbf{p})
+{\epsilon}_{2}^{0}{n}_{2}(\mathbf{p})
+\frac{1}{2} U(n_{1}(\mathbf{p})-\frac{1}{2})^{2}
+\frac{1}{2} U(n_{2}(\mathbf{p})-\frac{1}{2})^{2}
+\frac{1}{2} U_{m}({n}_{1}(\mathbf{p})-\frac{1}{2})({n}_{2}(\mathbf{p})-\frac{1}{2})}\,,
\end{equation} 
where $n_{1}(\mathbf{p})$ and $n_{2}(\mathbf{p})]$ are distribution functions for two species of fermions ("partilces" and "holes"). For the flat bands induced by the repulsive interaction, we have:
\begin{equation}
{\epsilon}_{1}=\frac{{\delta}E}{{\delta} n_{1}(\mathbf{p})}=0 \,\,,\,\,{\epsilon}_{2}=\frac{{\delta}E}{{\delta} n_{2}(\mathbf{p})}=0\,,
\label{e12}
\end{equation}
which give
\begin{eqnarray}
{\epsilon}_{1}^{0}+U(n_{1}(\mathbf{p})-\frac{1}{2})+\frac{1}{2}U_{m}(n_{2}(\mathbf{p})-\frac{1}{2})=0\,,
\\
{\epsilon}_{2}^{0}+U(n_{2}(\mathbf{p})-\frac{1}{2})+\frac{1}{2}U_{m}(n_{1}(\mathbf{p})-\frac{1}{2})=0\,.
\end{eqnarray}
The distribution functions of particles and hole are:
\begin{equation}
n_{1}(\mathbf{p})=\frac{
4U^{2}-U_{m}^{2}-8U{\epsilon}_{1}^{0}+4U_{m}{\epsilon}_{2}^{0}}{8U^{2}-2U_{m}^{2}},
\end{equation}
and
\begin{equation}
n_{2}(\mathbf{p})=\frac{
4U^{2}-U_{m}^{2}-8U{\epsilon}_{2}^{0}+4U_{m}{\epsilon}_{1}^{0}}{8U^{2}-2U_{m}^{2}}.
\end{equation}
Those regions within which $0<n_{1}(\mathbf{p})<1$ and $0<n_{2}(\mathbf{p})<1$ are the flat bands in momentum space, as shown in the corresponding flat-band configuration. With increasing $|\mathbf{p}^{0}|$, several Lifshitz transitions occur at the critical points of Lifshitz given by:
\begin{equation}
p_{1}=p_{F}-\frac{2U-U_{m}}{4c},
\label{1T}
\end{equation}
\begin{equation}
p_{2}=\sqrt{p_{F}^{2}-\frac{2mU+mU_{m}}{2}},
\label{2T}
\end{equation}
\begin{equation}
p_{T}=\sqrt{p_{F}^{2}+\frac{2mU+mU_{m}}{2}},
\label{HT}
\end{equation}
\begin{equation}
p_{3}=p_{F}+\frac{2U-U_{m}}{4c},
\label{3T}
\end{equation}
\begin{equation}
p_{D}=\frac{2c^{2}m^{2}(2U+U_{m})^{2}+(U_{m}-2U)^{2}[2p_{F}^{2}+m(2U+U_{m})]}{4mc(4U^{2}-U_{m}^{2})}.
\label{D}
\end{equation}
At these transitions the Fermi bands appear, disappear or touch each other with formation of singular configurations.

\section{Results and Discussion}
\label{sec:Results}

We present theoretical results obtained from the effective Hamiltonians introduced in the previous sections. The purpose of these calculations is not to claim material-specific experimental fitting, but to demonstrate, in a direct and quantitative way, that the topological scenarios discussed in this paper generate identifiable spectral signatures associated with Lifshitz reconstruction, emergent horizon behaviour, transport of Berry monopoles, and enhanced flat-band pairing scales. The numerical plots should therefore be understood as normalized model calculations. They are intended to clarify the spectral geometry and topology of the effective Hamiltonians, rather than to reproduce the band structure of a particular experimental compound.

The results are organized around three related levels of description. The first level is the local tilted Weyl cone, where the control parameter directly tunes the spectrum from type-I to type-II form. This provides the cleanest mathematical picture of the Lifshitz transition because the zero-energy structure can be read directly from the quasiparticle branches. The second level is the horizon model, where the tilt is no longer introduced as an abstract parameter but is generated by the Painlev\'e--Gullstrand frame-drag velocity. This connects the type-I/type-II transition with the black-hole horizon analogy. The third level is the displaced Weyl model, where the central object is not only the overtilting of a cone but the motion of a Berry monopole relative to inner and outer Fermi surfaces. This allows the topological transfer process to be visualized explicitly.

These three levels should not be treated as disconnected examples. They express the same underlying idea from different perspectives. In each case, the important object is the zero-energy manifold. When this manifold changes from a point, to a line-like or critical structure, to an extended Fermi contour or pocket, the system undergoes a Lifshitz reconstruction. The transition is not primarily a symmetry-breaking event. It is a reorganization of the low-energy spectral topology.

We first consider the simple tilted Weyl Hamiltonian introduced in Eq.~(\ref{HamiltonianSimple}),
\begin{equation}
H=c{\mbox{\boldmath$\sigma$}}\cdot\hat{\mathbf{p}}-fcp_z \, .
\end{equation}
The corresponding quasiparticle energies are
\begin{equation}
E_{\pm}(\mathbf{p})=-fcp_z \pm c|\mathbf{p}| \, .
\end{equation}
The zero-energy manifold is determined by
\begin{equation}
-fp_z \pm |\mathbf{p}|=0 \, .
\end{equation}
This equation is the central diagnostic for the type-I/type-II transition. For small tilt, the equation can only be satisfied at the Weyl point. For critical tilt, the zero-energy condition becomes singular in the direction selected by the tilt. For overtilt, finite zero-energy contours appear. Thus the parameter \(f\) does not merely deform the cone visually; it changes the topology of the Fermi-level set.

For $f<1$, the tilt is not strong enough to generate a finite zero-energy contour. The spectrum retains its type-I Weyl form, and the Fermi surface remains point-like at the Weyl node. In this regime the Weyl point is still the only zero-energy solution. The cone is tilted, but it is not tilted enough for the upper or lower branch to cross the zero-energy plane away from the node. This is the ordinary Weyl-semimetal situation, where the low-energy physics is controlled by a point-like topological defect in momentum space.

At the critical value $f=1$, the cone reaches the boundary between the type-I and type-II regimes. In the simplified relativistic model, this critical condition corresponds to the Dirac-line configuration discussed analytically earlier. The critical state is therefore not a trivial midpoint between two cone shapes. It is the spectral boundary at which the zero-energy condition changes dimension. The determinant analysis in the preceding section shows that this critical regime is associated with the emergence of a nodal line and the corresponding \(N_2\) topological invariant. This is why the transition is naturally interpreted as a Lifshitz transition rather than as a smooth geometric deformation of the same phase.

For $f>1$, the tilt exceeds the critical value and the system enters the type-II regime. The Weyl cone becomes overtilted, and finite zero-energy contours appear. These contours represent the emergence of electron-like and hole-like sectors that meet at the Weyl node. The appearance of finite zero-energy structure is the clearest spectral signature of the Lifshitz reconstruction. The Weyl point remains present, but it is no longer the only object at the Fermi level. It becomes the contact point between extended Fermi structures.

The anisotropy of the zero-energy structure is also important. The finite contours do not appear isotropically in momentum space. They are organized around the direction selected by the tilt term, here the \(p_z\) direction. This means that the Lifshitz reconstruction has a directional character. The system does not simply acquire a larger Fermi surface; rather, the topology of the zero-energy manifold is reorganized around a preferred axis. This directional feature becomes central in the horizon analogy, where the preferred direction is supplied by the radial frame-drag flow.

\begin{figure}[t]
\centering
\includegraphics[width=0.9\linewidth]{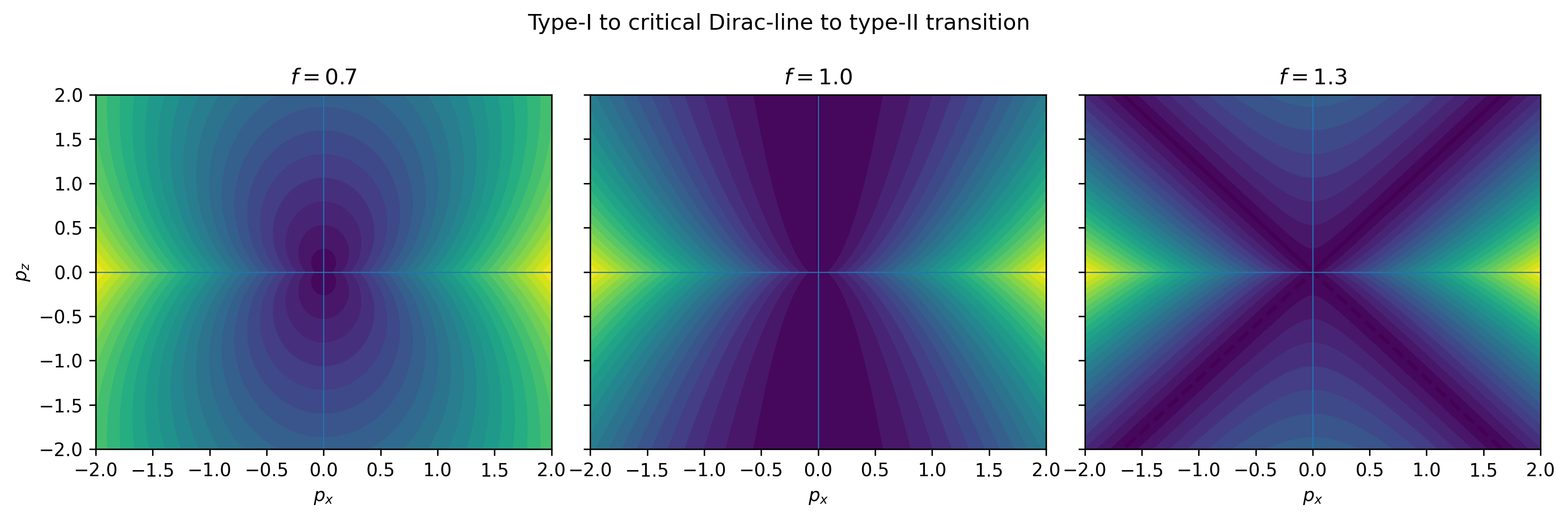}
\caption{Evolution of the zero-energy structure of the tilted Weyl Hamiltonian across the Lifshitz transition. For $f<1$, the zero-energy set collapses to the Weyl point. At $f=1$, the critical Dirac-line condition is reached. For $f>1$, finite zero-energy contours appear, corresponding to the type-II Weyl regime.}
\label{fig:tilted_transition_sequence}
\end{figure}

Figure~\ref{fig:tilted_transition_sequence} shows the reconstruction across three regimes. For $f<1$, the zero-energy structure is point-like. At $f=1$, the system reaches the critical Dirac-line condition. For $f>1$, the zero-energy set opens into finite contours. This sequence is useful because it shows that the type-II regime does not arise abruptly as an unrelated structure. It emerges through a well-defined critical configuration at the Lifshitz boundary. Both quasiparticle branches are included, because the type-II Fermi structure is determined by the zero-energy loci of \(E_+\) and \(E_-\) together. A single lower-band energy map may show how one spectral branch bends under overtilting, but the complete type-II Fermi structure requires both branches. The physical interpretation of type-II Weyl behaviour depends on the coexistence and contact of electron-like and hole-like sectors. The structure confirms that the overtilted regime cannot be described simply as a more strongly slanted type-I cone. The zero-energy set has changed its character. In the type-I case, the Weyl point is isolated and the low-energy excitation spectrum is locally conical around that point. In the type-II case, the zero-energy plane cuts through the tilted cone in such a way that an extended set of momenta becomes available at zero energy. 

The sequence in Fig.~\ref{fig:tilted_transition_sequence} gives the simplest numerical confirmation of the analytical classification. In the subcritical regime, the Fermi-level topology is that of an isolated point. At criticality, the zero set becomes singular and the nodal-line description becomes relevant. In the supercritical regime, the system supports extended zero-energy contours. The transition is therefore visible without needing to calculate a material-specific observable. The zero-energy geometry alone already shows the change in phase structure.

The Dirac line appears at the critical boundary between type-I and type-II behaviour in the simplified relativistic model, mediatING the change in codimension of the zero-energy set. The presence of the \(N_2\) invariant in the analytical discussion is therefore directly connected to the numerical reconstruction shown in the figure.

We next evaluate the horizon Hamiltonian introduced in Eq.~(\ref{HamiltonianBH}),
\begin{equation}
H=\pm c{\mbox{\boldmath$\sigma$}}\cdot{\bf p}-p_r v(r)+\frac{c^2p^2}{E_{\rm UV}} \, ,
\end{equation}
where the radial frame-drag velocity is
\begin{equation}
v(r)=c\sqrt{\frac{r_h}{r}} \, .
\end{equation}
The second term acts as a Doppler shift generated by the Painlev\'e--Gullstrand frame-drag velocity, while the quadratic term provides ultraviolet regularization. This Hamiltonian turns the abstract tilt parameter into a spatially varying effective velocity. The horizon is then identified with the location where this velocity reaches the speed \(c\).

The horizon condition is determined by \(v(r)=c\), which occurs at \(r=r_h\). Outside the horizon, where \(r/r_h>1\), one has \(v(r)<c\), and the Weyl cone remains tilted but not overtilted. At the horizon, where \(r/r_h=1\), the tilt reaches the critical value. Inside the horizon, where \(r/r_h<1\), one has \(v(r)>c\), and the Weyl spectrum enters the type-II regime. This gives a direct spectral interpretation of the horizon: it is the spatial surface on which the Weyl spectrum crosses the Lifshitz boundary.

The linear frame-drag term drives the cone into the type-II regime. The quadratic term regularizes the high-momentum behaviour and closes the resulting Fermi-pocket structure. Without such a term, the effective low-energy model would show overtilted behaviour but would not by itself provide a physically bounded momentum-space pocket. The regularization is not a numerical convenience but expresses the fact that the low-energy Weyl description must eventually be completed in higher-energy physics regime.

In the normalized sequence used here, the representative positions are \(r/r_h=1.05\), \(r/r_h=1.00\), and \(r/r_h=0.95\). These correspond to outside, at, and inside the horizon. For the interior value \(r/r_h=0.95\), the velocity ratio is
\begin{equation}
\frac{v}{c}=\sqrt{\frac{1}{0.95}}\approx1.026 \, .
\end{equation}
This corrected value should be used consistently in the figure caption and discussion. The earlier value \(v/c=1.1\) would correspond to a deeper interior point, approximately \(r/r_h=0.826\), not \(r/r_h=0.95\).

\begin{figure}[t]
\centering
\includegraphics[width=0.55\linewidth]{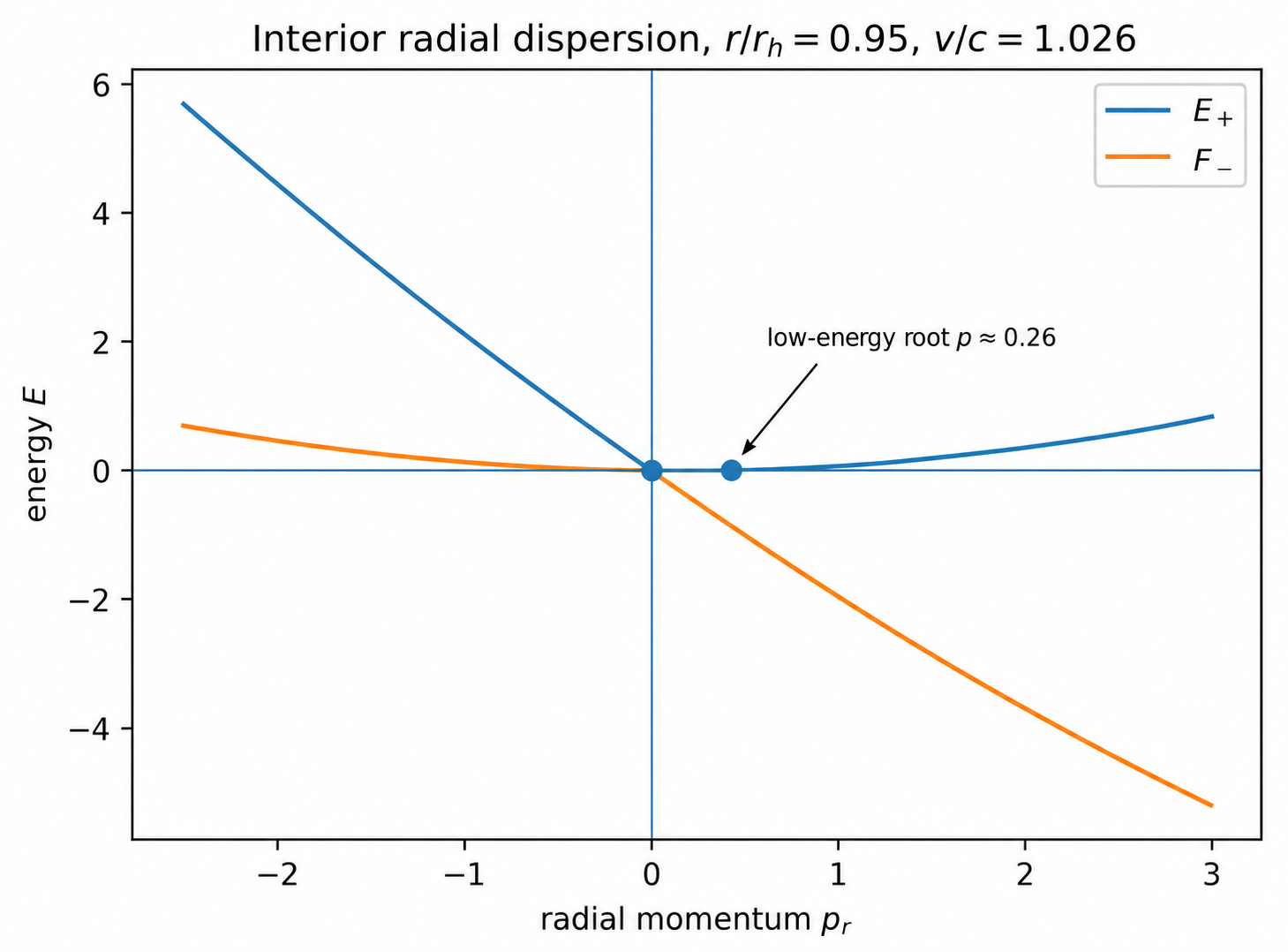}
\caption{Radial dispersion for the Painlev\'e--Gullstrand Weyl Hamiltonian in the interior region at $r/r_h=0.95$, where $v/c=\sqrt{1/0.95}\approx1.026$. The overtilted spectrum develops an additional zero-energy crossing, indicating the appearance of bounded interior Fermi-pocket structure.}
\label{fig:radial_inside}
\end{figure}

Figure~\ref{fig:radial_inside} shows the radial dispersion in the interior region at \(r/r_h=0.95\). The overtilted spectrum develops an additional zero-energy crossing away from the original Weyl point. This new crossing is the local spectral signature that the system has moved into the type-II regime. In physical terms, it indicates that additional zero-energy states have become available inside the horizon.

The additional crossing in Fig.~\ref{fig:radial_inside} should not be interpreted as a minor perturbative feature, as the spectral evidence that the interior region has a different Fermi-level topology from the exterior region. Outside the horizon, the spectrum remains type-I and the zero-energy structure is tied to the original Weyl point. Inside the horizon, the effective tilt exceeds the critical value and additional zero-energy solutions appear. This is exactly the type of reconstruction expected from a Lifshitz transition.

In the Painlev\'e--Gullstrand representation, the frame-drag velocity enters the Weyl Hamiltonian in the same structural role as the tilt term in the simple Weyl model. Thus crossing the horizon is mathematically analogous to increasing the tilt parameter through its critical value. The emergent type-II spectrum behind the horizon is therefore not imposed by hand. It follows from the same overtilting mechanism already displayed in the simple model.

\begin{figure}[t]
\centering
\includegraphics[width=0.55\linewidth]{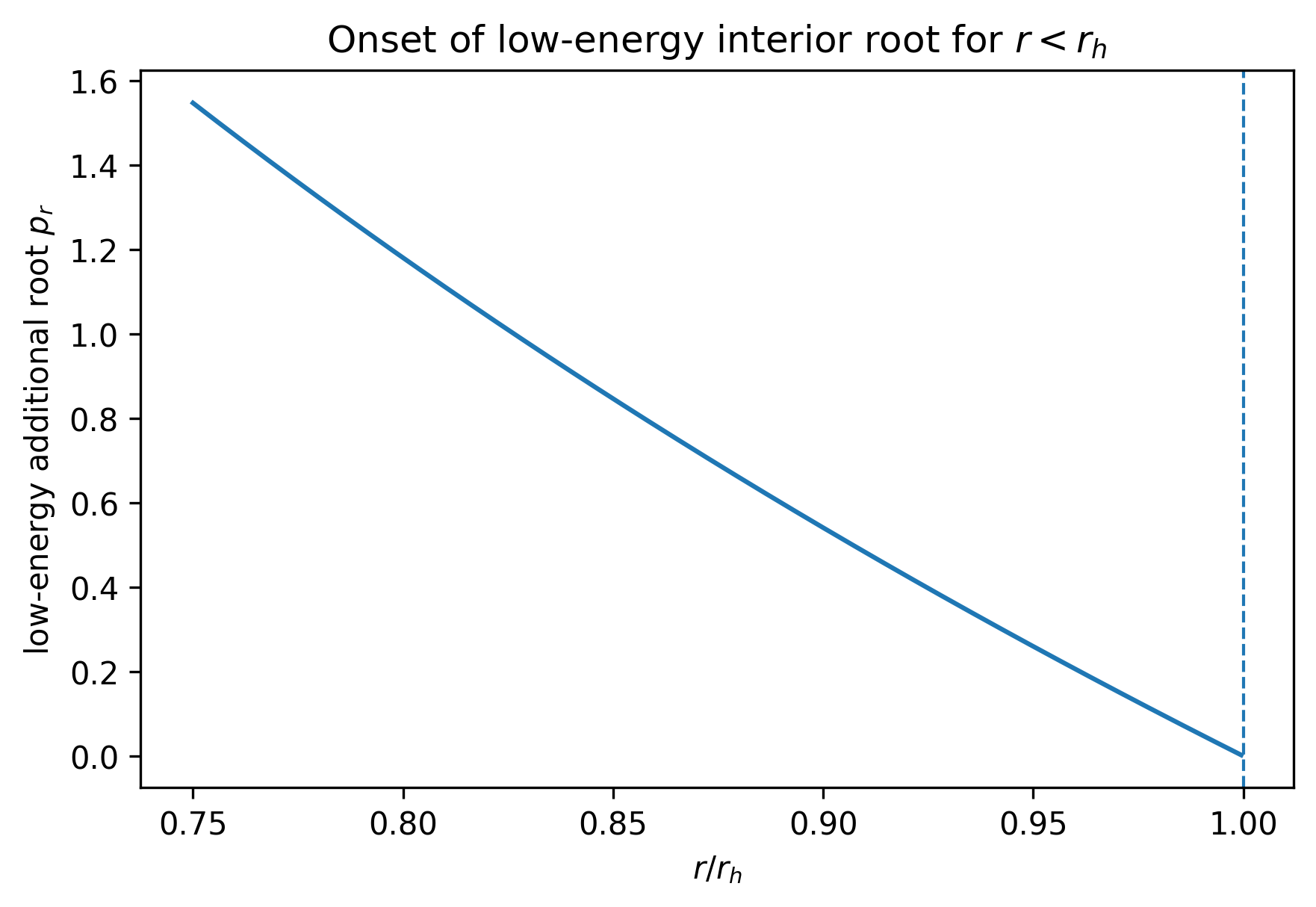}
\caption{Appearance of the additional zero-energy root across the horizon. In the normalized horizon model, the non-trivial root appears only for $r/r_h<1$, where $v(r)>c$, and grows continuously inside the overtilted region.}
\label{fig:zero_roots}
\end{figure}

Figure~\ref{fig:zero_roots} tracks the non-trivial zero-energy root across the horizon. The additional root is absent outside the horizon, appears at \(r/r_h=1\), and grows only in the interior region where \(v(r)>c\). This root-tracking view complements the radial-dispersion plot. The dispersion plot shows the local shape of the branches at one interior position, while the root plot shows the onset of the new zero-energy structure across the spatial horizon.

The behaviour in Fig.~\ref{fig:zero_roots} gives a clean diagnostic of horizon crossing in spectral terms. The horizon is not only the position where \(v=c\). It is also the onset point for the additional zero-energy solution. This makes the Lifshitz interpretation precise: the transition occurs when the number and topology of zero-energy solutions changes. The fact that the additional root grows continuously inside the horizon shows that the interior Fermi-pocket structure develops smoothly from the critical point, rather than appearing as an arbitrary discontinuity.

This root structure also supports the discussion of non-equilibrium filling. The calculation itself does not model the kinetics of occupation or Hawking emission. However, it identifies the spectral precondition for such processes: after crossing the horizon, new zero-energy states exist in the interior region. The filling of these states is then the physical process that may be associated with analogue Hawking radiation in the semimetal interpretation.

The horizon analogy may also be expressed through the effective Hawking-temperature scale. For the Painlev\'e--Gullstrand velocity field,
\begin{equation}
v(r)=c\sqrt{\frac{r_h}{r}} \, ,
\end{equation}
the derivative at the horizon is negative. However, temperature is a positive quantity, so the magnitude of the derivative must be used:
\begin{equation}
T_{{\rm H}}=\frac{\hbar}{2\pi}\left|\frac{dv}{dr}\right|_{r=r_h}
=\frac{\hbar c}{4\pi r_h} \, .
\end{equation}
In the normalized units used in the plot, \(\hbar=c=k_B=1\), this becomes
\begin{equation}
T_{{\rm H}}=\frac{1}{4\pi r_h} \, .
\end{equation}

This temperature scale should be interpreted carefully. The plot does not claim that a particular laboratory semimetal has the exact thermal scale shown. Instead, it shows the normalized inverse dependence of the analogue Hawking scale on the horizon radius. The point is conceptual and scaling-based: sharper or smaller effective horizons correspond to larger analogue temperature scales, while larger horizons correspond to smaller scales. This is the expected behaviour because the effective gravitational field is controlled by the spatial gradient of the frame-drag velocity at the horizon.

\begin{figure}[t]
\centering
\includegraphics[width=0.55\linewidth]{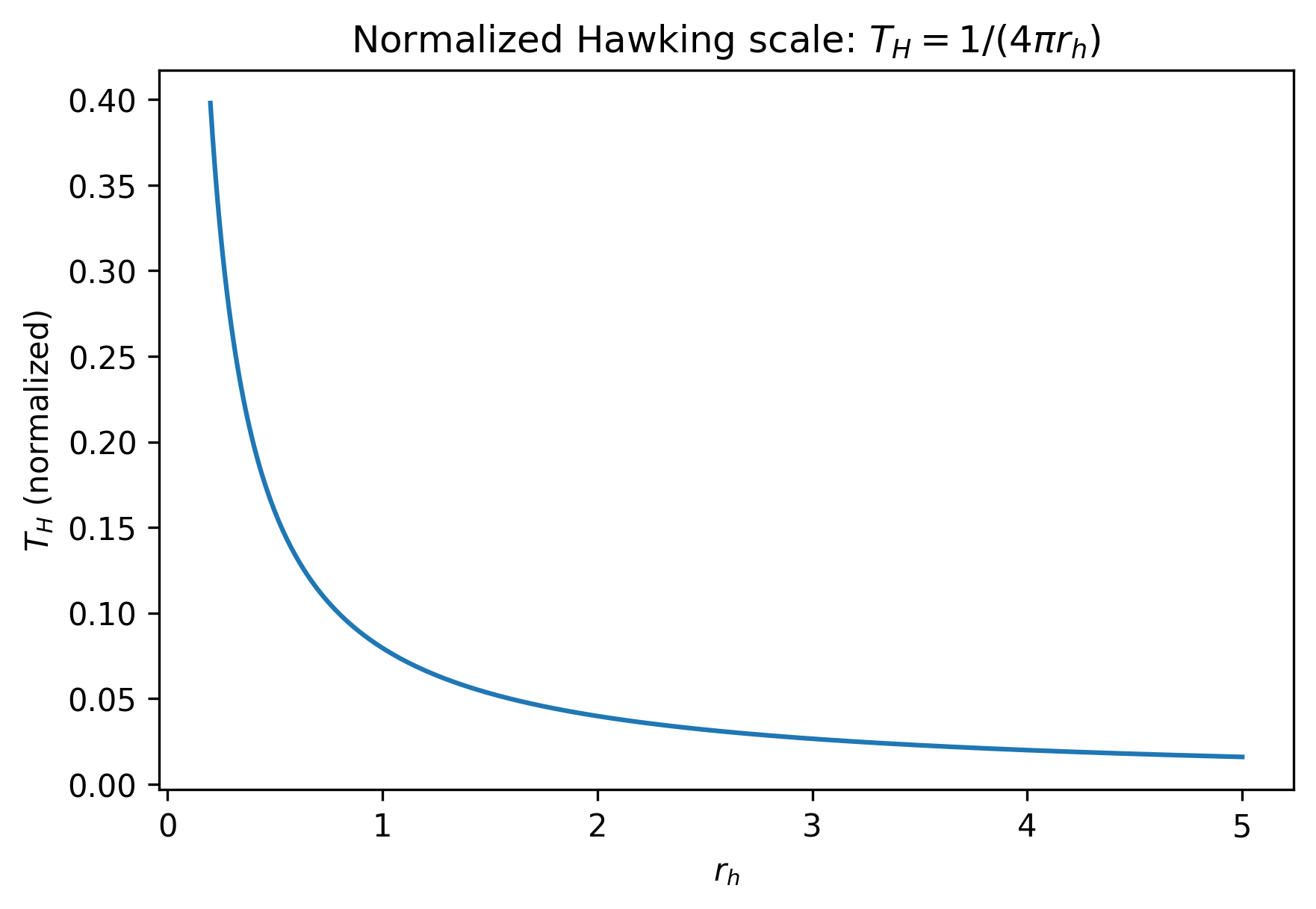}
\caption{Normalized Hawking-temperature scale as a function of the horizon radius $r_h$ in units with $\hbar=c=k_B=1$. The scale follows $T_H=1/(4\pi r_h)$, showing that smaller effective horizons correspond to larger analogue Hawking temperatures.}
\label{fig:hawking_radius}
\end{figure}

Figure~\ref{fig:hawking_radius} connects the spectral horizon picture with a physically interpretable scale. The previous figures show where the zero-energy structure changes; this figure shows how the same horizon geometry determines the effective thermal scale associated with the analogue process. The dependence is inverse in \(r_h\), so the thermal scale is controlled by the sharpness of the velocity profile at the horizon. This is consistent with the general principle that analogue Hawking effects are governed by gradients at the horizon rather than by the absolute value of the velocity far from the horizon.

The use of the magnitude \(\left|dv/dr\right|\) is essential. If the derivative is used without the absolute value, the expression would produce a negative value for a black-hole velocity profile, which is not physically meaningful for a temperature. The sign of the derivative reflects the direction of the flow variation, not the sign of the thermal scale. Therefore, the corrected equation and figure use the positive magnitude.

We now turn to the displaced Weyl Hamiltonian introduced in Eq.~(\ref{ExampleHamiltonian}),
\begin{equation}
H=c{\mbox{\boldmath$\sigma$}}\cdot(\mathbf{p}-\mathbf{p}^{(0)})+\frac{p^2-p_F^2}{2m} \, .
\end{equation}
This model describes a Weyl point displaced by \(\mathbf{p}^{(0)}\) relative to a background Fermi-surface scale \(p_F\). The displacement parameter \( |\mathbf{p}^{(0)}| \) controls the position of the Berry monopole relative to the inner and outer Fermi surfaces. The model therefore allows the topological rearrangement to be discussed not only as a change in spectral shape but also as a transport process involving Berry flux.

For the numerical evaluation, we use the normalized parameter set
\begin{equation}
m=1,\qquad c=1,\qquad p_F=2 \, .
\end{equation}
This choice lies in the regime \(p_F>mc\), as assumed in the analytical discussion. Along the displacement direction, the quasiparticle energies may be written as
\begin{equation}
E_{\pm}(p)=\frac{p^2-p_F^2}{2m}\pm c|p-p^{(0)}| \, .
\end{equation}
This one-dimensional cut is not the full momentum-space geometry, but it is useful because it shows the critical touching events clearly.

The first transition occurs at
\begin{equation}
|\mathbf{p}^{(0)}|=p_F=2 \, .
\end{equation}
At this value, the Weyl node touches the Fermi surface. For \( |\mathbf{p}^{(0)}|<p_F \), the inner and outer Fermi surfaces are arranged so that they both enclose the Weyl point. At the critical value, the Weyl point lies exactly on the Fermi surface. This produces the type-II Weyl contact configuration that mediates the topological rearrangement.

The second transition occurs at
\begin{equation}
|\mathbf{p}^{(0)}|=\frac{m^2c^2+p_F^2}{2mc}=2.5 \, .
\end{equation}
This is the point at which the inner Fermi surface collapses and disappears. These two transitions have different meanings. The first is a touching and flux-transfer event. The second is the disappearance of a Fermi surface after it is no longer protected by the same topological configuration. The distinction is important because it shows that the displaced Weyl model contains a genuine two-stage Lifshitz sequence.

\begin{figure}[t]
\centering
\includegraphics[width=0.55\linewidth]{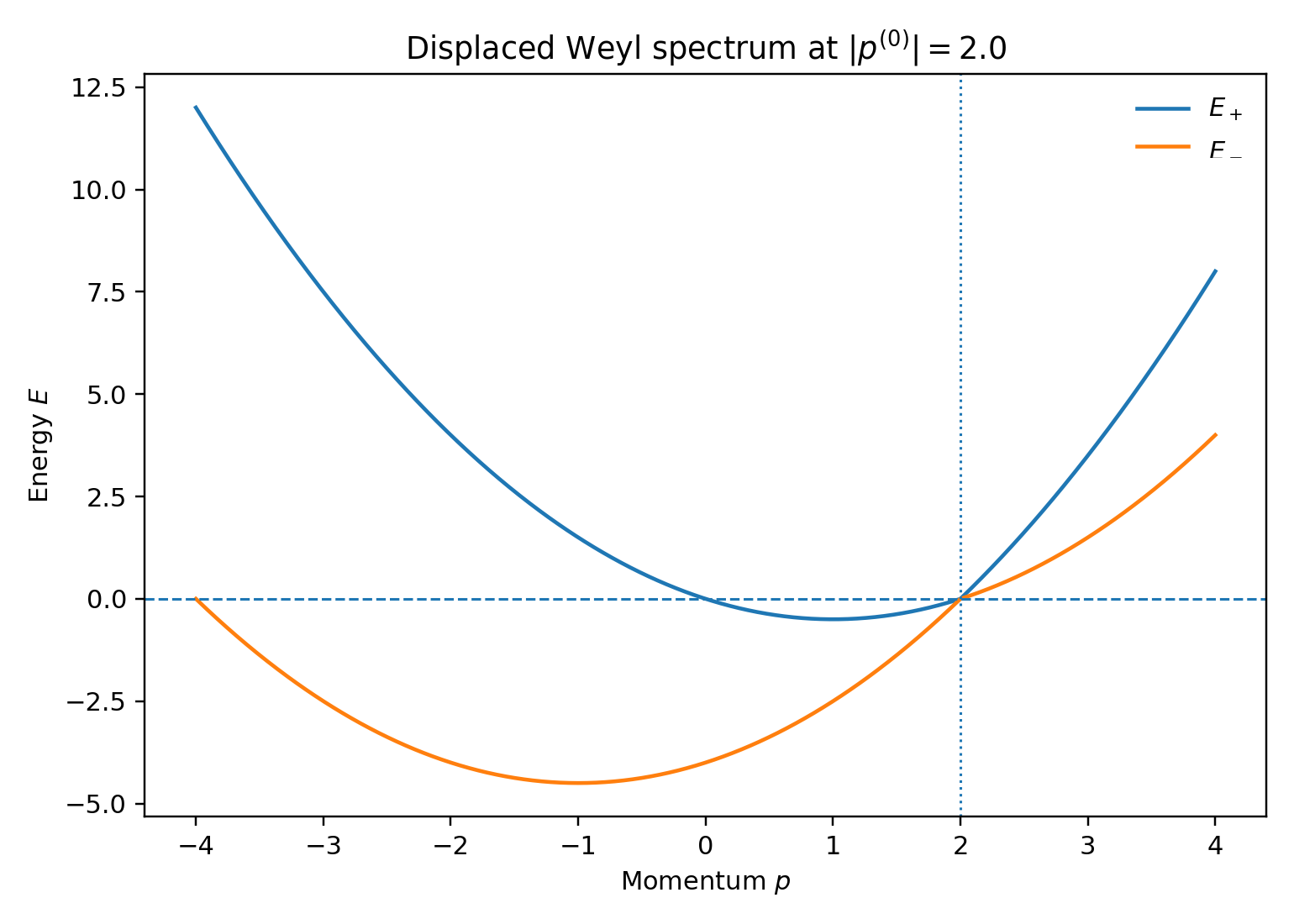}
\caption{Displaced Weyl spectrum along the displacement axis at $|\mathbf{p}^{(0)}|=2.0$, corresponding to the first Lifshitz transition for the normalized choice $m=1$, $c=1$, and $p_F=2$. At this point the Weyl node touches the Fermi surface and mediates the topological reconstruction.}
\label{fig:displaced_line_p20}
\end{figure}

Figure~\ref{fig:displaced_line_p20} shows the spectrum along the displacement axis at the first critical value. The touching is visible directly in the line spectrum and gives a local microscopic view of the topological reconstruction. This figure is particularly useful because it connects the compact analytical condition \( |\mathbf{p}^{(0)}|=p_F \) with the actual spectral crossing at the Weyl node.

The spectral contact in Fig.~\ref{fig:displaced_line_p20} is the local signature of a global topological process. The Weyl point carries Berry monopole charge, and its motion relative to the Fermi surfaces changes which surfaces enclose that charge. At the first transition, the Weyl point is precisely at the contact point. This is why the transition is not merely a crossing of ordinary bands. It is a topological exchange event in which the relationship between the monopole and the surrounding Fermi surfaces is reorganized.

This interpretation also explains why the displaced Weyl model adds something not already contained in the simple tilted-cone model. The simple tilted model shows how type-II structure appears from overtilting. The displaced model shows how a Weyl node can move through Fermi surfaces and change their global topological content. Both are Lifshitz phenomena, but they highlight different aspects of the same class of transitions.

\begin{figure}[t]
\centering
\includegraphics[width=0.60\linewidth]{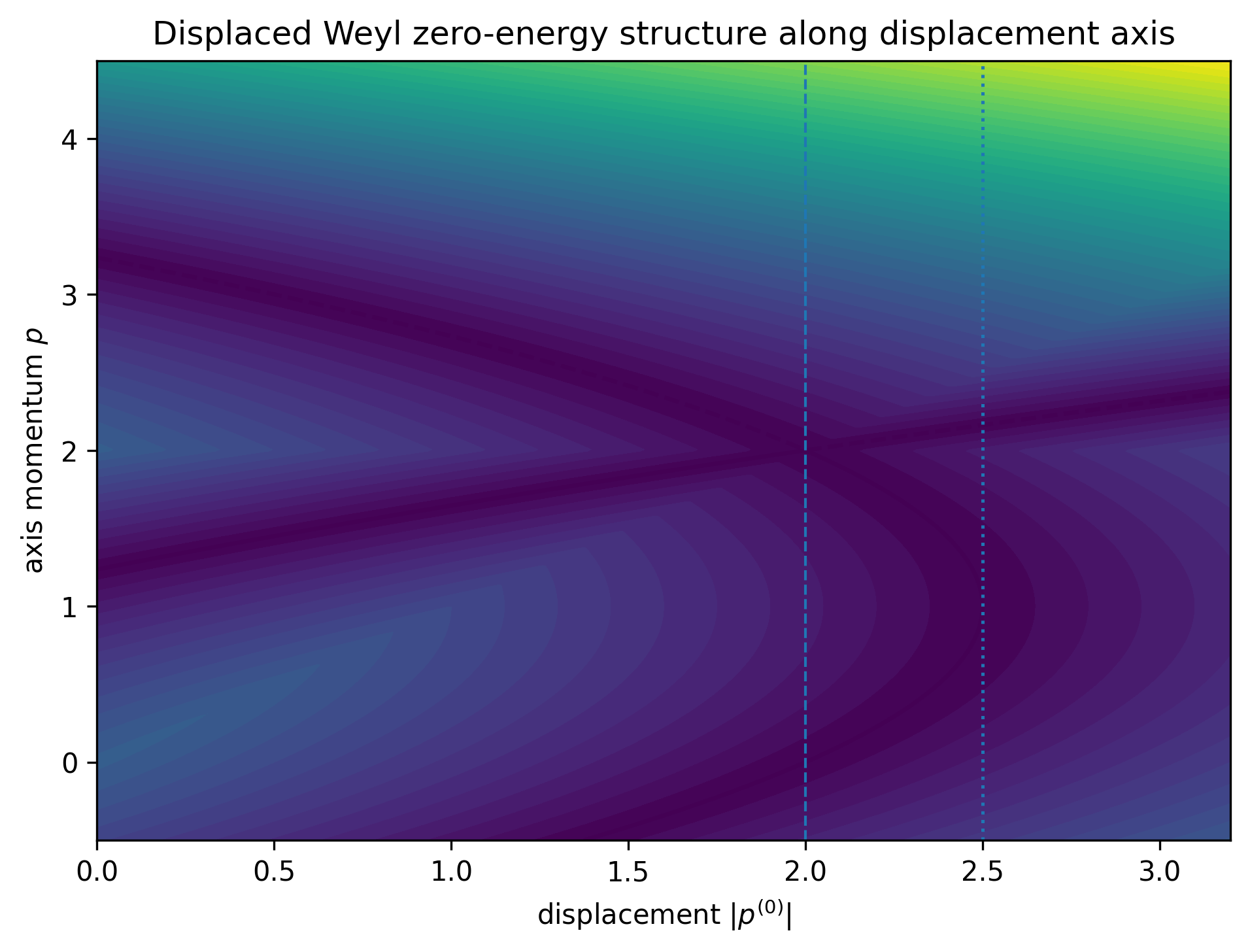}
\caption{Transition map for the displaced Weyl model in the regime $p_F>mc$. The zero-energy structure changes as the displacement $|\mathbf{p}^{(0)}|$ is varied. The dashed and dotted horizontal lines mark the first transition at $|\mathbf{p}^{(0)}|=p_F$ and the second transition at $|\mathbf{p}^{(0)}|=(m^2c^2+p_F^2)/(2mc)$, respectively.}
\label{fig:displaced_map}
\end{figure}

Figure~\ref{fig:displaced_map} compresses the full displacement-driven evolution into a single transition map. The zero-energy structure changes as \( |\mathbf{p}^{(0)}| \) is varied, and the two marked critical values separate the first touching event from the later collapse of the inner surface.

The transition map in Fig.~\ref{fig:displaced_map} shows that the intermediate region between the two critical values has its own physical meaning. After the first transition, the Weyl point has crossed the Fermi surface configuration, but the inner surface has not yet disappeared. This intermediate phase therefore represents a reorganized Fermi-surface topology rather than an immediate collapse. Only at the second transition does the inner surface vanish. The map makes this staged process visible.

The regime condition \(p_F>mc\) should be retained in the caption and discussion because the ordering and interpretation of the two critical scales depend on it. In the normalized example \(p_F=2\), \(m=1\), and \(c=1\), this condition is clearly satisfied. Under this condition, the first transition occurs at \(2\), and the second occurs at \(2.5\). The numerical figure therefore agrees with the analytical expressions quoted in the theory section.

\begin{figure}[t]
\centering
\includegraphics[width=0.60\linewidth]{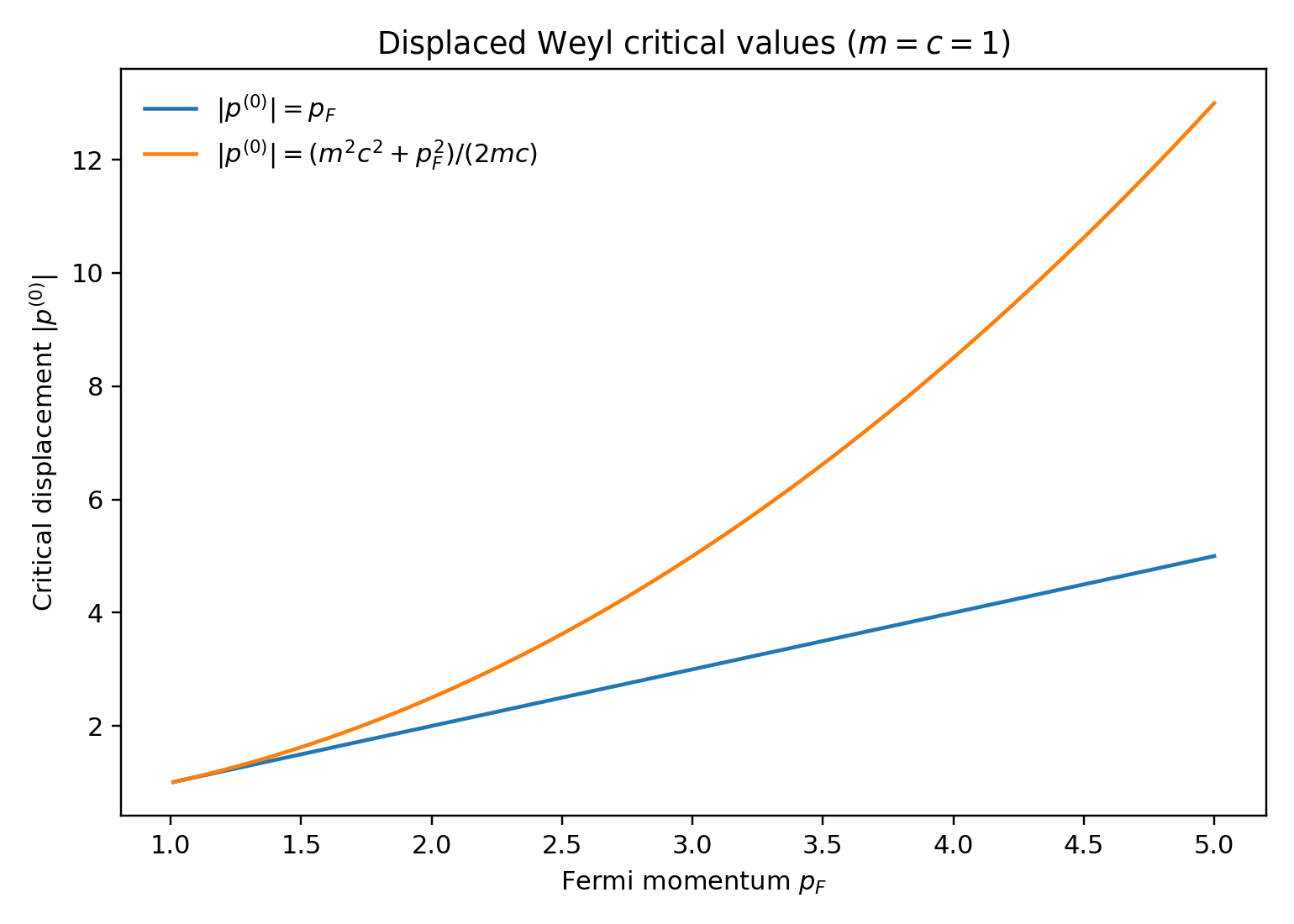}
\caption{Critical displacement values in the displaced Weyl model as functions of $p_F$ in the regime $p_F>mc$. The first critical value is $p_F$, while the second is $(m^2c^2+p_F^2)/(2mc)$. Their separation shows that the Fermi-surface touching event and the collapse of the inner surface are distinct Lifshitz transitions.}
\label{fig:critical_values_displaced}
\end{figure}

The broader dependence of these two critical values on \(p_F\) is shown in Fig.~\ref{fig:critical_values_displaced}. The first critical value grows linearly with \(p_F\), while the second follows
\begin{equation}
|\mathbf{p}^{(0)}|=\frac{m^2c^2+p_F^2}{2mc} \, .
\end{equation}
This comparison shows that the two transition scales separate generically in the regime considered here. Their separation demonstrates that the touching event and the disappearance of the inner surface are structurally distinct.

Figure~\ref{fig:critical_values_displaced} places the normalized example into a wider parameter context. It shows that the choice \(p_F=2\) is not a special accidental case. The double-transition structure is a general feature of the model in the stated regime. The linear branch identifies the first contact between the Weyl point and the Fermi surface, while the nonlinear branch identifies the later collapse of the inner surface. This reinforces the interpretation that the displaced Weyl model contains two distinct topological events.

The separation between the two curves can also be read as a measure of the size of the intermediate regime. When the curves are far apart, the system spends a larger range of displacement values in the reorganized phase between first contact and inner-surface collapse. When the curves approach one another, the intermediate regime narrows. Thus the critical-values plot is not merely a mathematical summary. It indicates how the parameter space of the model is partitioned into different spectral-topological regimes.

The final set of results concerns the flat-band pairing enhancement discussed in connection with Eq.~(\ref{expVSlin}). The analytical discussion shows that interaction effects near Lifshitz-critical configurations may produce nearly dispersionless states with singular density of states. This changes the qualitative dependence of the pairing scale on the interaction strength. In a conventional metallic state, the gap scale has the exponentially suppressed form
\begin{equation}
\Delta_{\rm normal}=E_0 \exp\left(-\frac{1}{gN_F}\right) \, ,
\end{equation}
whereas in the flat-band case the scale is linear in the coupling,
\begin{equation}
\Delta_{\rm flat\,band}=\frac{gV_d}{2(2\pi\hbar)^d} \, .
\end{equation}

The distinction between these two expressions is central to the physical significance of flat-band formation. In the normal case, weak coupling produces an exponentially small gap. This is why ordinary pairing instabilities may remain strongly suppressed unless the coupling or density of states is sufficiently favourable. In the flat-band case, the dispersionless structure produces a singular density of states and changes the dependence from exponential to linear. This does not automatically predict room-temperature superconductivity in a specific material, but it shows why Lifshitz-critical flat-band formation can dramatically strengthen pairing tendencies.

We evaluate both expressions only in normalized form. The normal-state gap is computed with \(E_0=1\) and \(N_F=1\), while the flat-band expression is represented by a linear coefficient chosen for clarity of comparison. 

\begin{figure}[t]
\centering
\includegraphics[width=0.60\linewidth]{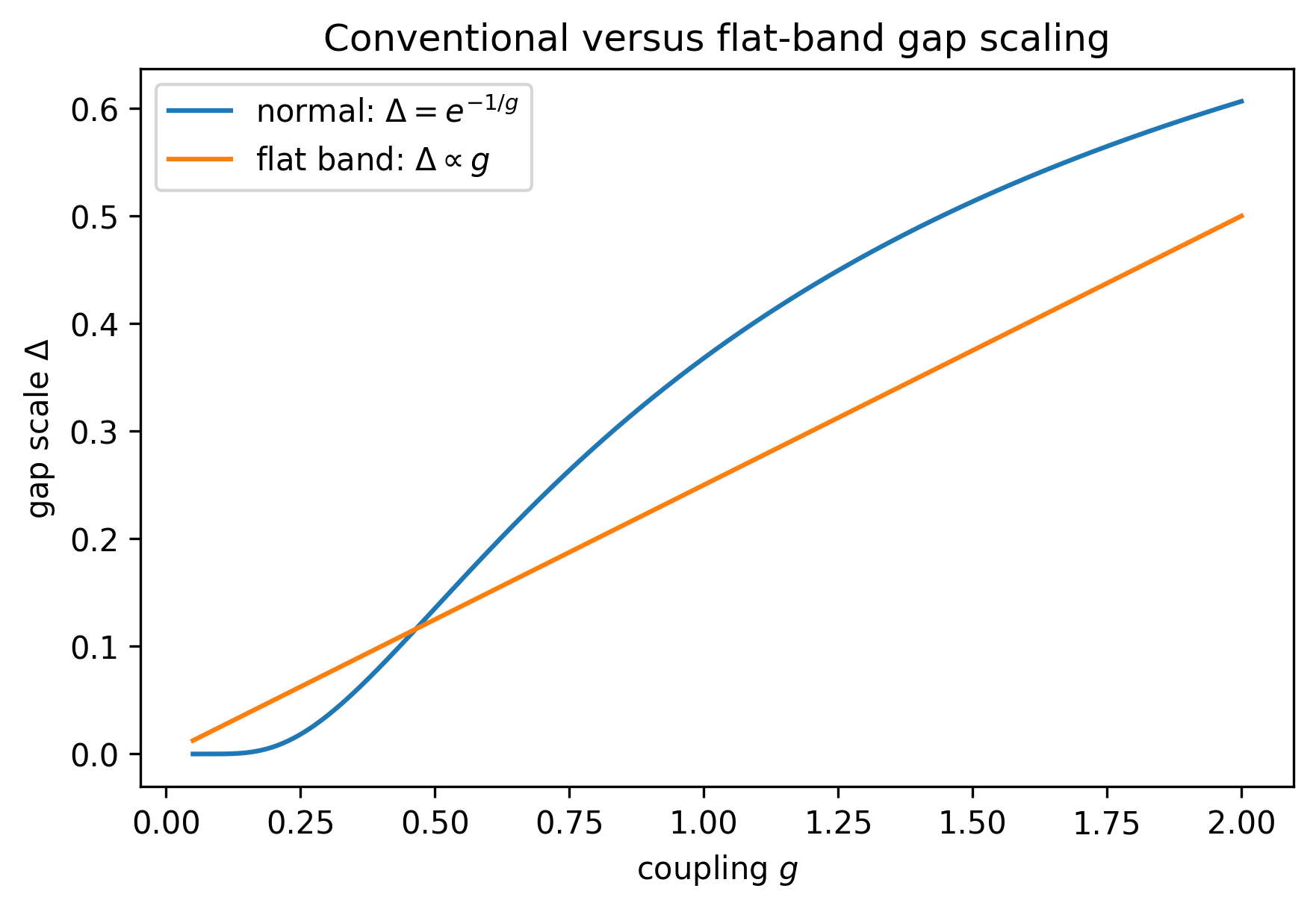}
\caption{Normalized comparison between the conventional exponentially suppressed pairing scale and the linear flat-band pairing scale. The figure is intended as a scaling illustration rather than a material-specific prediction.}
\label{fig:gap_scaling}
\end{figure}

Figure~\ref{fig:gap_scaling} illustrates why flat-band formation is physically important in the present discussion. The conventional curve remains small over a broad weak-coupling range because the dependence is exponential. The flat-band curve grows linearly from the onset of coupling. In the context of this paper, this means that the Lifshitz transition is not only a topological or geometric rearrangement of the Fermi surface. Once interactions are included, it may also strongly affect the scale of many-body instabilities.

This observation links the flat-band discussion back to the earlier Weyl and horizon calculations. The earlier figures show how zero-energy structures appear and reorganize. The flat-band figure shows why such zero-energy structures can matter dynamically. If the spectral reconstruction produces nearly dispersionless or high-density low-energy states, then interaction effects may become much more prominent. Thus the topological Lifshitz transition can act as a gateway to enhanced many-body response.

\begin{figure}[t]
\centering
\includegraphics[width=0.60\linewidth]{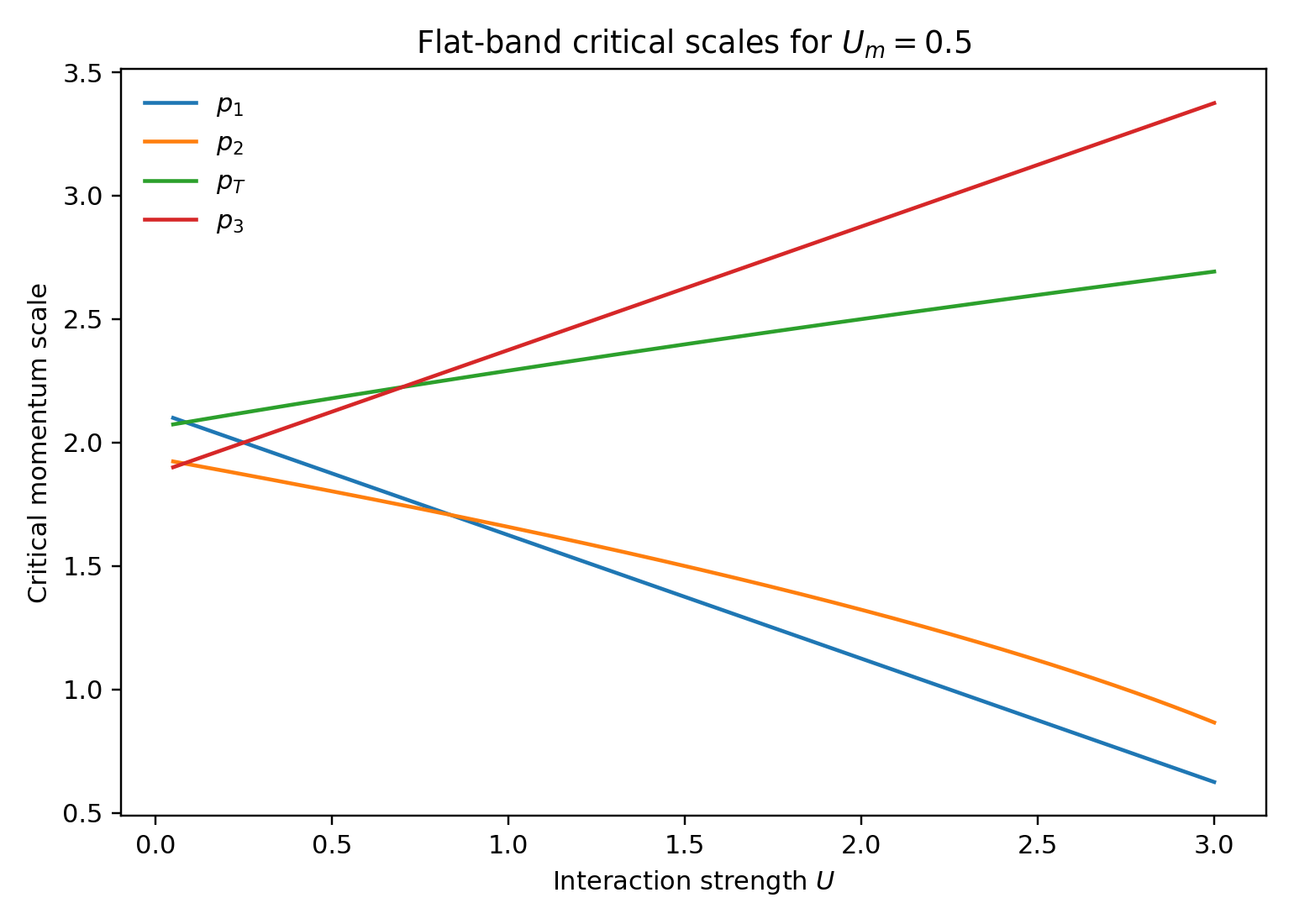}
\caption{Flat-band critical scales as functions of the interaction strength $U$ for fixed $U_m=0.5$. The multiple branches show that the interacting flat-band sector contains several distinct critical momenta associated with separate Lifshitz reconstructions.}
\label{fig:flatband_scales}
\end{figure}

To go beyond the simple gap comparison, we also retain the interaction-dependent critical scales in the flat-band sector. Figure~\ref{fig:flatband_scales} plots the critical momenta \(p_1\), \(p_2\), \(p_T\), and \(p_3\) as functions of the interaction strength for fixed \(U_m=0.5\). These scales come from the interaction-dependent flat-band construction introduced earlier. They describe the momentum-space boundaries at which partially occupied regions appear, disappear, or touch.

Figure~\ref{fig:flatband_scales} shows that the interacting flat-band problem is not governed by a single threshold. Several critical scales evolve with interaction strength, which means that the flat-band sector contains its own internal Lifshitz structure. This prevents the interaction discussion from being reduced to a simple statement that interactions increase the gap. The interactions reorganize the allowed partially occupied regions in momentum space, and these reorganizations have their own critical boundaries. The different branches move at different rates as \(U\) increases. This indicates that the flat-band regions do not expand or contract uniformly. Some boundaries shift outward, others shift inward, and some may only exist over restricted parameter intervals depending on the square-root conditions in the analytical expressions. The figure therefore demonstrates that interaction-driven flat-band formation is a structured topological process, not merely a smooth energetic correction.

The simple tilted Weyl model exhibits a clear spectral transition at \(f=1\), with the zero-energy structure changing from point-like type-I form to extended type-II form through a critical Dirac-line condition. The Painlev\'e--Gullstrand Weyl Hamiltonian shows that crossing the event horizon corresponds spectrally to crossing the same type-I to type-II boundary, with an additional zero-energy root appearing only inside the horizon. The displaced-monopole Hamiltonian reproduces a double Lifshitz sequence in which a Weyl touching event transfers topological structure between Fermi surfaces and is followed by the disappearance of the inner surface at a second critical value. The gap-scaling and flat-band critical-scale plots show how Lifshitz-critical flat-band physics can produce strong many-body consequences when interaction effects are included.

The type-I to type-II Weyl transition, the horizon analogy, the transport of Berry monopoles, and the flat-band enhancement mechanism are different manifestations of the same underlying principle. When the zero-energy structure of a Weyl system is reorganized through a Lifshitz transition, both its topology and its low-energy spectral response may change qualitatively. The calculations identify the spectral and topological conditions under which such analogies and many-body consequences become plausible. The value of the model calculations is that they isolate the relevant mechanisms: overtilting, zero-energy reconstruction, monopole transport, interior root formation, and flat-band scaling. These mechanisms provide a consistent theoretical bridge between Weyl Lifshitz transitions, emergent horizon analogies, and interaction-enhanced flat-band physics.

\section{Conclusions}

We discussed the transitions which involve the nodes of different co-dimensions: the Fermi surfaces with topological
charge $N_{1}$ (co-dimension 1), Weyl points with the topological charge $N_{3}$ (co-dimension 3)
and Dirac lines with topological charge $N_{2}$ (co-dimension 2). Depending on the
type of transition, the intermediate state has the type-II Dirac
point, the type-II Weyl point or the Dirac line. The latter is supported by combination of symmetry
and topology. There are different configurations of the Fermi surfaces, involved in the  Lifshitz transition with the Weyl points in the intermediate state. The interplay of different topological invariants enhances the variety
of the topological Lifshitz transitons. 
In the corresponding type-II Weyl schematic the type-II Weyl point connects
the Fermi pockets, and the Lifshitz transition corresponds to the transfer of the Berry flux between the Fermi pockets.
In the corresponding second type-II Weyl schematic the type-II Weyl point connects the outer and inner Fermi surfaces. At the Lifshitz transition the Weyl point is released from both Fermi surfaces. They lose their Berry flux and the 
topological charge $N_3$, which guarantees the global stability. As a result the inner surface disappears after shrinking to a point at the second Lifshitz transition.

Many other Lifshitz transitions are expected, since
we did not touch here the other possible topological features: topological
invariants which describe the shape of the Fermi surface; the shape
of the Dirac nodal lines; their interconnections; etc. The interplay
of topologies can be seen in particular in the electronic spectrum
of Bernal and rombohedral graphite.\cite{mcclure57,Mikitik2006,Mikitik2008,HeikkilaVolovik2015}
In particular, in the electronic spectrum of Bernal graphite the type-II
Dirac line has been identified, which is connected with the type-I
Dirac line at some point in the 3D momentum space.\cite{HyartHeikkilaVolovik2016}
If one considers $p_{z}$ as parameter, then at some critical value
of $p_{z}$ there is the transition from the 2D type-I Dirac point
to the 2D type-II Dirac point.

However, the most important property of Lifshitz transitions is that in the vicinity of the topological transition the electron-electron interaction leads to the formation of zeros in the spectrum of the co-dimension 0, i.e. to the flat bands. Because of the singular density of electronic states, materials with the flat band are the plausible candidates 
for room-temperature superconductivity.

\section*{Author Contributions}

Iftekher S. Chowdhury: numerical simulations, methodology and model implementation

Hom Nath Dhungana: mathematical analysis, modelling and simulation validation

Binay Prakash Akhouri: theoretical discussion and physical interpretation

Shah Haque: scholarly discussion, results presentation, and manuscript review.

Hind Adawi: scholarly discussion, theoretical review, feedback, and results validation

Eric Howard: conceptualization, methodology, theoretical framing, manuscript preparation, and final review.

All authors have read and agreed to the submitted version of the manuscript.

\section*{Funding}
This research received no external funding.

\section*{Data Availability Statement}
No experimental datasets were generated or analyzed during the current study. The results are based on analytical model Hamiltonians and normalized numerical simulations described in the manuscript. The simulation code and generated figures can be made available by the corresponding author upon reasonable request.

\section*{Acknowledgements}
The authors acknowledge their respective institutions for academic support.

\section*{Conflicts of Interest}
The authors declare no conflict of interest.

\end{document}